\documentclass[twocolumn]{aastex631}
\usepackage{newtxtext,newtxmath}
\shorttitle{Oxygen isotope fractionation in the Martian atmosphere}
\shortauthors{Yoshida et al.}
\graphicspath{{./}{figures/}}

\begin{document}

\title{Oxygen isotope fractionation in the Martian atmosphere induced by CO$_2$ photolysis and O$_3$ formation}

\author{Tatsuya Yoshida}
\affiliation{Earth-Life Science Institute, Institute of Science Tokyo, Tokyo, Japan}
\affiliation{Department of Earth and Space Sciences, University of Washington, Seattle, WA, USA}
\affiliation{Department of Geophysics, Graduate School of Science, Tohoku University, Miyagi, Japan}

\author{Shohei Aoki}
\affiliation{Department of Complexity Science and Engineering, Graduate School of Frontier Sciences, The University of Tokyo, Chiba, Japan}
\affiliation{Department of Geophysics, Graduate School of Science, Tohoku University, Miyagi, Japan}

\author{Hiromu Nakagawa}
\affiliation{Department of Geophysics, Graduate School of Science, Tohoku University, Miyagi, Japan}

\author{Naoki Terada}
\affiliation{Department of Geophysics, Graduate School of Science, Tohoku University, Miyagi, Japan}

\author{Juan Alday}
\affiliation{Instituto de Astrof\'{i}sica de Andaluc\'{i}a, CSIC, Granada, Spain}

\author{Akinori Hasebe}
\affiliation{Department of Geophysics, Graduate School of Science, Tohoku University, Miyagi, Japan}

\author{Yuki Nakamura}
\affiliation{Research Center for Advanced Science and Technology, The University of Tokyo, Tokyo, Japan}

\author{Shungo Koyama}
\affiliation{Department of Earth and Planetary Sciences, School of Science, Institute of Science Tokyo, Japan}
\affiliation{Department of Geophysics, Graduate School of Science, Tohoku University, Miyagi, Japan}

\author{Shotaro Sakai}
\affiliation{Faculty of Environment and Information Studies, Keio University, Kanagawa, Japan}

\author{Ryoya Sakata}
\affiliation{Research Center for Advanced Science and Technology, The University of Tokyo, Tokyo, Japan}

\author{Ann Carine Vandaele}
\affiliation{Royal Belgian Institute for Space Aeronomy, Brussels, Belgium}
\affiliation{Department of Geophysics, Graduate School of Science, Tohoku University, Miyagi, Japan}



\begin{abstract}
	The enrichment of heavy isotopes of volatile elements in the Martian atmosphere indicates that Mars lost a large portion of its atmosphere through escape to space. Recent atmospheric measurements by ExoMars Trace Gas Orbiter (TGO) have suggested that the vertical profiles of oxygen isotopic compositions are influenced by chemical reactions involving isotopic fractionation. However, their quantitative impacts have not yet been fully evaluated. In this study, we develop a 1D photochemical model that incorporates oxygen isotopic fractionation associated with CO$_2$ photolysis and O$_3$ formation to investigate the vertical profiles of oxygen isotopic compositions. Our calculations show that CO is depleted in heavy oxygen isotopes relative to CO$_2$, reaching $\delta ^{18}$O $\sim -25$‰ and $\delta ^{17}$O $\sim -15$‰, primarily due to isotopic fractionation during CO$_2$ photolysis. The vertical profiles of oxygen and carbon isotopic compositions are in good agreement between our model and the TGO measurements. O$_3$ is strongly enriched in $^{18}$O and $^{17}$O, reaching $\delta ^{18}$O $\sim 100$‰ and $\delta ^{17}$O $\sim 50$‰ as a consequence of the isotopic fractionation during its formation, whereas atomic oxygen is highly depleted in the heavy oxygen isotopes with $\delta ^{18}$O $\lesssim -100$‰ and $\delta ^{17}$O $\lesssim -50$‰ so as to compensate for their enrichment in O$_3$. These chemical fractionation processes can deplete the heavy oxygen isotopes in species that escape from the upper atmosphere, and thereby enhance the isotopic fractionation associated with oxygen escape to space. Such fractionated isotopic compositions of escaping oxygen may be detectable by the Martian Moons eXploration (MMX) mission.
\end{abstract}



\section{Introduction} \label{sec:intro}
	The Martian atmosphere is known to be enriched in heavy isotopes of volatile elements such as hydrogen, carbon, nitrogen, and noble gases relative to Earth and primitive meteorites \citep[e.g.,][]{Owen1977,Webster2013}. This enrichment indicates that Mars lost a large portion of its atmosphere through atmospheric escape processes. Therefore, the isotopic composition has been widely used as a tracer of the atmospheric evolution of Mars \citep[e.g.,][]{Jakosky1991,Pepin1991,Jakosky1994,Pepin1994,Hu2015,Kurokawa2018,Hu2022,Thomas2023,Lyons2024}. 
	
	The isotopic fractionation associated with atmospheric escape is controlled not only by the escape mechanism itself, but also by the vertical distribution of isotopic compositions in the atmosphere, which is primarily shaped by chemical reactions involving isotopic fractionation and vertical transport \citep{Lyons2024}. Recent atmospheric measurements by ExoMars Trace Gas Orbiter (TGO) have revealed that the vertical profiles of carbon and oxygen isotopic compositions are strongly influenced by atmospheric chemical processes. \citet{Alday2023} and \citet{Aoki2023} retrieved the vertical profile of the carbon and oxygen isotopic compositions of CO, which is a major photochemical product of CO$_2$, from solar occultation measurements performed by TGO. They found that CO is depleted in $^{13}$C relative to CO$_2$ by more than $\sim 100$‰. The pronounced $^{13}$C depletion can be explained by isotopic fractionation associated with CO$_2$ photolysis, which arises from the smaller UV absorption cross section of $^{13}$CO$_2$ compared to $^{12}$CO$_2$ \citep{Schmidt2013,Ueno2024}. The role of CO$_2$ photolysis in shaping the carbon isotopic composition is further supported by comparisons with atmospheric photochemical models that incorporate isotopic fractionation \citep{Alday2023,Yoshida2023}.
	
	The TGO measurements also suggest that the oxygen isotopic composition of CO is fractionated compared to CO$_2$, while the magnitude of oxygen fractionation is estimated to be smaller than that of carbon \citep{Alday2023,Aoki2023}. CO$_2$ photolysis can induce oxygen isotopic fractionation owing to differences in UV absorption cross sections among O-bearing isotopologues \citep{Schmidt2013}. In addition, ozone formation is known to generate strong mass-independent fractionation in the Earth's stratosphere \citep[e.g.,][]{Thiemens1983,Mauersberger1987,Brinjikji2021}, leading to enrichments of $^{18}$O and $^{17}$O in ozone through the preferential formation of asymmetric isotopologues such as OO$^{18}$O and OO$^{17}$O \citep[e.g.,][]{Gao2001,Ivanov2013}. This isotopic fractionation in ozone can be transferred to other O-bearing species through subsequent oxygen exchange reactions \citep{Yung1991,Lyons2001,Liang2007}. Although these oxygen isotopic fractionation processes are expected to strongly influence the isotopic compositions of O-bearing species in the Martian atmosphere and can potentially explain the oxygen isotopic fractionation suggested by the TGO measurements, their quantitative impacts have not yet been fully evaluated.
	
	In this study, we develop a 1D atmospheric photochemical model that explicitly incorporates the oxygen isotopic fractionation associated with CO$_2$ photolysis and O$_3$ formation, in order to clarify their effects on the vertical profiles of oxygen isotopic compositions in the Martian atmosphere. The paper is organized as follows. In Section 2, we describe the outline of our photochemical model. In Section 3, we present the numerical results of the atmospheric profiles. In Section 4.1, we examine the parameter dependencies of the isotopic composition profiles. In Section 4.2, we compare our calculation results with the TGO measurements. In Section 4.3, we estimate the oxygen escape fractionation factor while accounting for the chemical isotopic fractionation. 

\begin{figure}[htbp]
	\centering
	\includegraphics[width=\columnwidth]{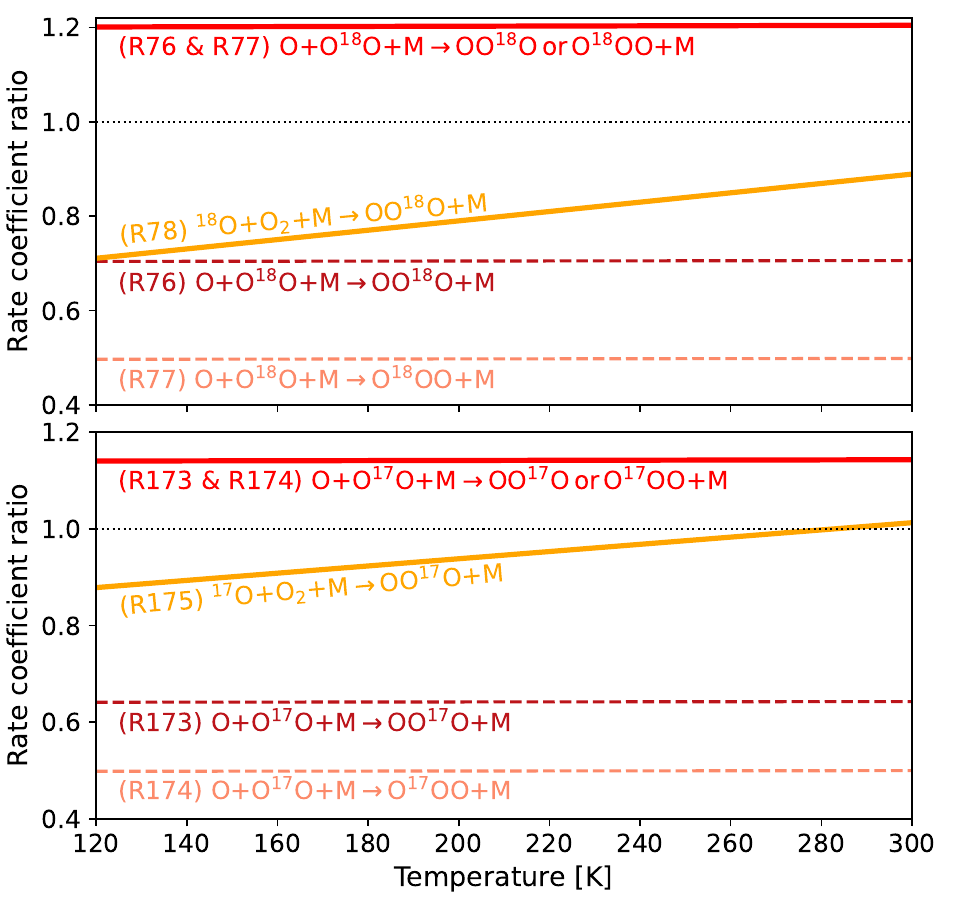}
	\caption{Rate coefficients of the formation reactions of $^{18}$O-bearing ozone (upper panel) and $^{17}$O-bearing ozone (lower panel) relative to that of O$_3$ ($=k_{\rm O+O_{2}+M}$) as a function of temperature. The solid red lines represent the rate coefficients of the reactions between O and O$^{18}$O (or O$^{17}$O), and the solid orange lines represent those of the reactions between $^{18}$O (or $^{17}$O) and O$_2$. The dashed lines represent the branching ratios of the production of symmetric and asymmetric ozone through the reactions between O and O$^{18}$O (or O$^{17}$O). The dotted black line represents the value corresponding to no isotopic fractionation.}
	\label{fig:1}
\end{figure}	

\begin{figure*}[htbp]
	\centering
	\includegraphics[width=2\columnwidth]{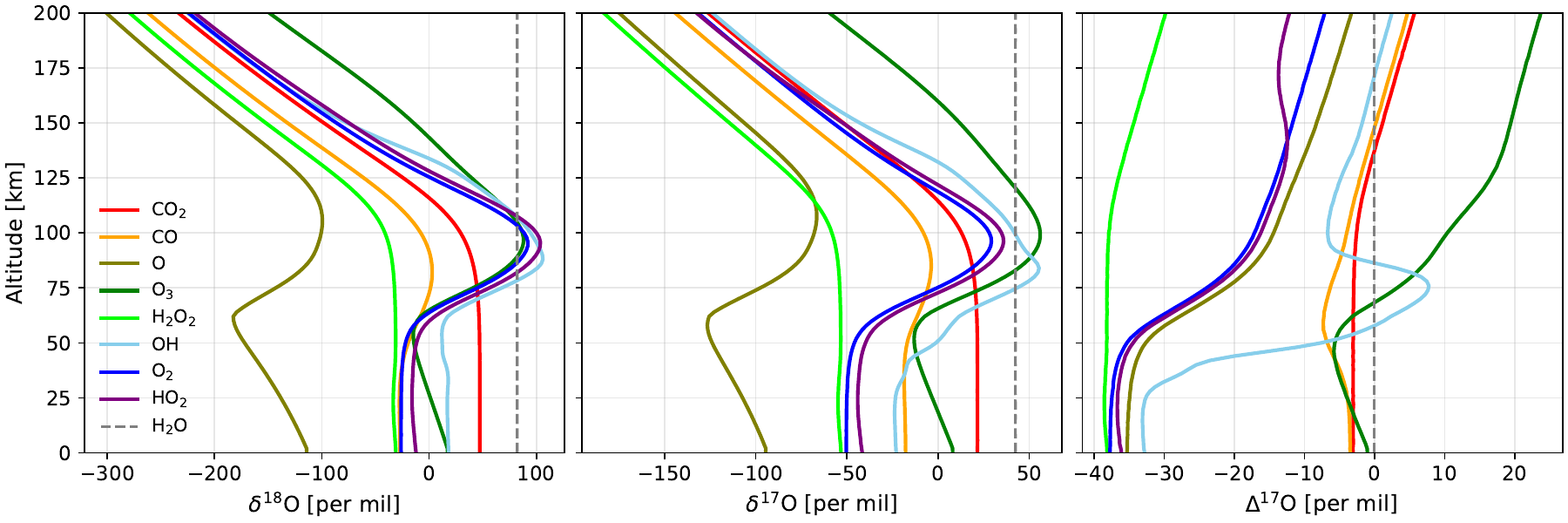}
	\caption{Vertical profiles of $\delta ^{18}$O (left), $\delta ^{17}$O (center), and $\Delta ^{17}$O (right) for each O-bearing species. The definitions of $\delta ^{j}$O ($j=17,\,18$) and $\Delta ^{17}$O are given in Equations (1) and (2), respectively. The dashed lines represent the fixed isotopic compositions of H$_2$O.}
	\label{fig:2}
\end{figure*}	

\section{Model description} \label{sec:model}
	We use our 1D photochemical model, PROTEUS \citep{Nakamura2023}, expanding the chemical network to include chemical species and reactions involving $^{18}$O and $^{17}$O. The overall model settings follow \citet{Yoshida2023}. Below, we describe the outline of the model and highlight the modifications introduced in this study relative to \citet{Yoshida2023}.
	
	We consider 245 chemical reactions for 40 species: CO$_2$, CO, O, O($^1$D), H$_2$O, H, OH, H$_2$, O$_3$, O$_2$, HO$_2$, H$_2$O$_2$, HOCO, CO$_{2}^{+}$, CO$^{18}$O, C$^{18}$O, $^{18}$O, $^{18}$O($^1$D), H$_2$$^{18}$O, $^{18}$OH, O$^{18}$OO, OO$^{18}$O, O$^{18}$O, HO$^{18}$O, H$_2$O$^{18}$O, HOC$^{18}$O, H$^{18}$OCO, CO$^{17}$O, C$^{17}$O, $^{17}$O, $^{17}$O($^1$D), H$_2$$^{17}$O, $^{17}$OH, O$^{17}$OO, OO$^{17}$O, O$^{17}$O, HO$^{17}$O, H$_2$O$^{17}$O, HOC$^{17}$O, and H$^{17}$OCO (Table A1). Here, we denote $^{16}$O simply as O. We explicitly distinguish symmetric and asymmetric ozone to consider the higher formation rate of the asymmetric one, which leads to enrichments of $^{18}$O and $^{17}$O in ozone (reactions R76--R78 and R173--R175 in Table A1). The reaction rate coefficients of the formation reactions of OO$^{18}$O, O$^{18}$OO, OO$^{17}$O, and O$^{17}$OO relative to that of O$_3$ are taken from \citet{Liu2021} and shown in Figure~\ref{fig:1}. Isotope exchange reactions between atomic oxygen and molecular oxygen, as well as those between atomic oxygen and carbon dioxide, are also taken into account (reactions R79, R80, R96, R97, R176, R177, R193, and R194 in Table A1) \citep{Gregory2021}. We additionally incorporate mass-dependent fractionation by scaling the reaction rate coefficients inversely with the square root of the reduced mass of the reactants, following \citet{Young2014}.
	
	We adopt the absorption cross sections of CO$_2$, CO$^{18}$O, and CO$^{17}$O at 138-212 nm provided by \citet{Schmidt2013} to compute the isotopic fractionation through CO$_2$ photolysis. At the other wavelengths, we apply the same absorption cross section as that of CO$_2$ provided by \citet{Huestis2011}. The absorption cross sections of the other $^{18}$O- and $^{17}$O-bearing species are assumed to be identical to those of their major isotopologues. For their absorption cross sections, we mainly refer to the JPL publication \citep{Burkholder2015} and the MPI-Mainz UV/VIS Spectral Atlas of Gaseous Molecules \citep{Keller2013}, as described in detail in \citet{Nakamura2023}. We apply the solar spectrum profile in the wavelength range from 0.5 to 1100 nm provided by \citet{Woods2009} to calculate the profiles of the photolysis rates. The spectral resolution is set to 0.1 nm in 138-212 nm to resolve the differences in the photolysis rate among CO$_2$, CO$^{18}$O, and CO$^{17}$O following \citet{Yoshida2023}, and 1 nm at all other wavelengths. The solar zenith angle is set to 0 degrees following \citet{Chaffin2017} and \citet{Yoshida2023}. Although it affects the optical depth and shifts the altitude at which the optical depth reaches unity, its impact on the atmospheric profiles is minor for non-extreme solar zenith angles (i.e., away from near-terminator conditions) compared to other parameters such as eddy diffusion. An example of the sensitivity test result for a solar zenith angle of 60 degrees is shown in Figure A5.
	
	Profiles of temperature, eddy diffusion coefficients, molecular diffusion coefficients, and H$_2$O number density are adopted from \citet{Yoshida2023} (Figures A1-A3). Here, the H$_2$O number density profile is constructed as follows. Below 30 km, the relative humidity is fixed at 22\%, corresponding to 9.5 precipitable microns of water. Above this level, the H$_2$O profile is connected to the saturation vapor mixing ratio at the altitude of the temperature minimum, and the same mixing ratio is assumed at higher altitudes \citep{Koyama2021,Yoshida2023}.
	
	The number density of CO$_2$ at the surface is fixed at $2.1\times 10^{17}\,\mathrm{cm^{-3}}$ \citep{Chaffin2017,Yoshida2023}. The oxygen isotopic ratio of CO$_2$ is also prescribed at the surface. Here, the oxygen isotopic ratio is expressed as
	\begin{equation}
		\delta ^{j}\mathrm{O}=\left(\frac{^{j}R}{^{j}R_s}-1\right)\times 1000,
	\end{equation}
	\begin{equation}
		\Delta ^{17}\mathrm{O}=\left[\mathrm{ln}\left(\frac{\delta ^{17}\mathrm{O}}{1000}+1\right) - 0.528\mathrm{ln}\left(\frac{\delta ^{18}\mathrm{O}}{1000}+1\right)\right]\times 1000,
	\end{equation}	
	where $j\,(=17,\,18)$ is the mass number of the minor oxygen isotopes, and $^{j}R$ and $^{j}R_{s}$ are the $^{j}$O/$^{16}$O ratio and its standard value, respectively. $\delta ^{j}\mathrm{O}$ and $\Delta ^{17}\mathrm{O}$ are expressed in per mil. $^{18}R_{s}$ and $^{17}R_{s}$ are the reference values of Vienna Standard Mean Ocean Water (VSMOW; $^{18}$O/$^{16}$O = $2.0052 \times 10^{-3}$, $^{17}$O/$^{16}$O = $3.7990 \times 10^{-4}$). $\delta ^{18}$O and $\delta ^{17}$O in CO$_2$ at the surface are set to 48‰ and 24‰, respectively, to reproduce the Curiosity measurements \citep{Webster2013}. The number density profiles of H$_{2}$$^{18}$O and H$_2$$^{17}$O are fixed to yield $\delta ^{18}$O=$84$‰ and $\delta ^{17}$O=$43$‰, respectively. Here, $\delta ^{18}$O of H$_2$O is taken from the Curiosity rover measurements \citep{Webster2013}. $\delta ^{17}$O is estimated from $\delta ^{18}$O assuming mass-dependent fractionation ($\Delta \mathrm{^{17}O}=0$) as no in situ measurements are currently available and TGO measurements near the surface have large uncertainties \citep{Alday2019}. At the upper boundary with an altitude of 200 km, H and H$_2$ are assumed to escape to space by Jeans escape, and the O escape rate is fixed at $1.2\times 10^{8}\,\mathrm{cm^{-2}}\,\mathrm{s^{-1}}$, following \citet{Chaffin2017} and \citet{Yoshida2023}. This assumption has no significant effect on the overall isotopic composition profile, as the timescale for atmospheric escape is much longer than that of the chemical reactions responsible for isotopic fractionation.

\section{Results} \label{sec:result}
\begin{figure}[htbp]
	\centering
	\includegraphics[width=\columnwidth]{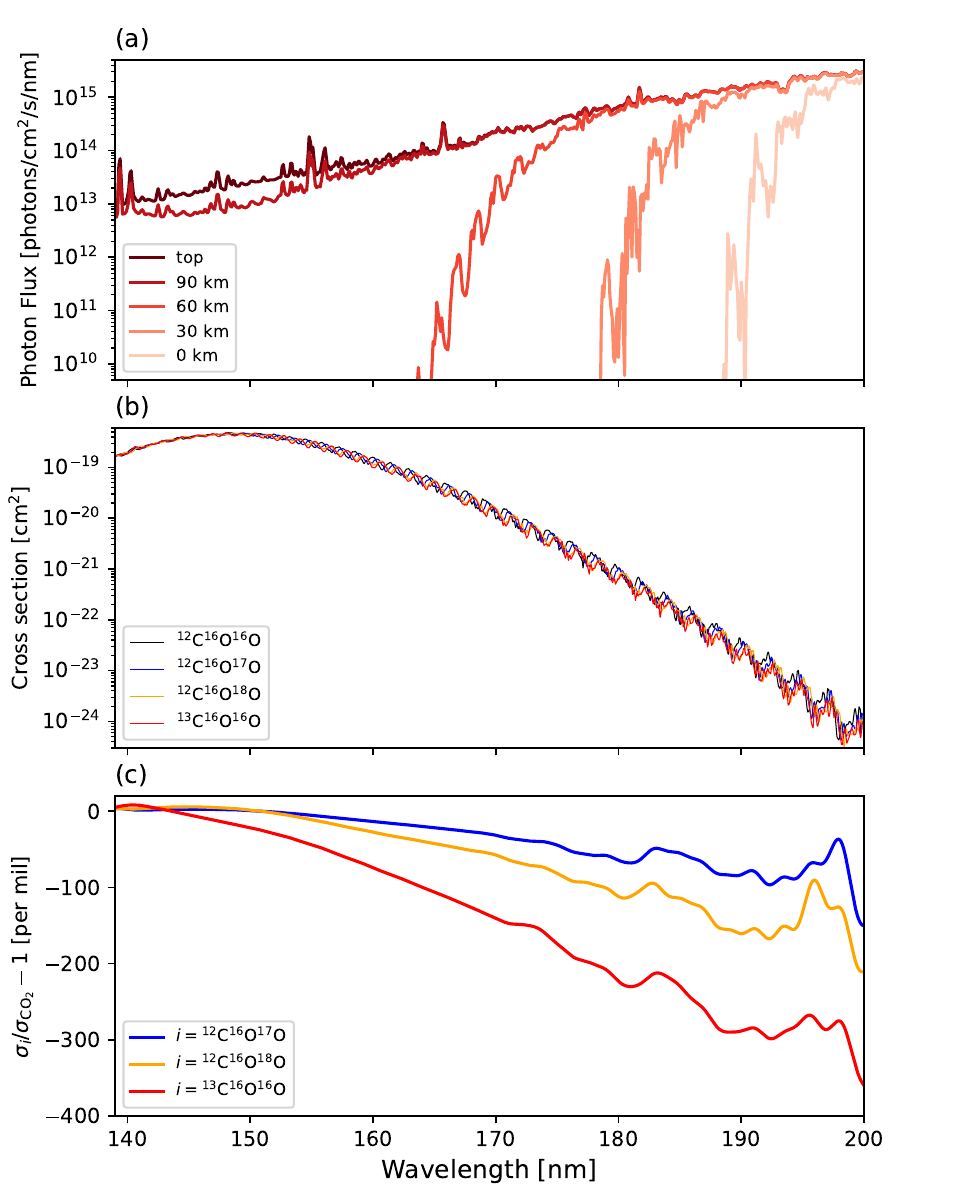}
	\caption{(a) Profile of photon flux with wavelength at selected altitudes. (b) Absorption cross sections of CO$_{2}$ (black), CO$^{17}$O (blue), CO$^{18}$O (orange), and $^{13}$CO$_2$ (red) with wavelength. (c) Relative difference in absorption cross section between species $i$ and CO$_2$, defined as $(\sigma_{i}/\sigma_{\rm CO_{2}}-1)\times 1000$, where $\sigma_{i}$ and $\sigma_{\rm CO_{2}}$ are the absorption cross sections of species $i$ and CO$_2$, respectively ($i=\mathrm{CO^{18}O}, \mathrm{CO^{17}O}, \mathrm{^{13}CO_{2}}$) \citep{Schmidt2013}. Relative differences are smoothed using a Gaussian window with FWHMs of 2.5 nm, as done by \citet{Schmidt2013}.}
	\label{fig:3}
\end{figure}

	The calculated vertical profiles of the oxygen isotopic compositions of each species are shown in Figure~\ref{fig:2}. The number density profiles are shown in Figure A4. With respect to the relationship between CO and CO$_2$, CO is depleted in $^{18}$O and $^{17}$O relative to CO$_2$ throughout the atmosphere (Figure~\ref{fig:2}). This depletion arises because CO is primarily produced by CO$_2$ photolysis, which induces oxygen isotopic fractionation due to the smaller absorption cross sections of CO$^{18}$O and CO$^{17}$O compared to CO$_2$ (Figures~\ref{fig:3}(b) and (c)). Below $\sim 100$ km, $\delta ^{18}$O and $\delta ^{17}$O of CO decrease as the altitude decreases. This reflects the shift of absorbed UV radiation toward longer wavelengths in the lower atmosphere (Figure~\ref{fig:3}(a)), where the differences in the absorption cross sections among isotopologues become more pronounced (Figures~\ref{fig:3}(b) and (c)). This behavior is analogous to that of the carbon isotopic profiles reported by \citet{Yoshida2023}, while the magnitude of the fractionation is moderate due to the smaller differences in the absorption cross sections (Figure~\ref{fig:3}(c)). Above $\sim 100\,\mathrm{km}$, corresponding to the homopause altitude, $\delta ^{18}$O and $\delta ^{17}$O of both CO and CO$_2$, as well as those of other species, decrease monotonically with increasing altitude (Figure~\ref{fig:2}) due to diffusive separation driven by molecular mass differences among isotopologues. $\Delta ^{17}$O of CO remains close to zero (Figure~\ref{fig:2}), reflecting the near-mass-dependent nature of isotopic fractionation associated with CO$_2$ photolysis \citep{Schmidt2013}. 
	
	O$_3$ tends to become enriched in $^{18}$O and $^{17}$O as a consequence of the isotopic fractionation during its formation. This enrichment becomes particularly pronounced above $\sim 60$  km. This is attributed to the increasing contribution of the ozone-forming reactions involving O and O$^{18}$O (or O$^{17}$O), which have large reaction rate coefficients relative to that of the O$_3$ formation reaction, with increasing mixing ratio of O (Figure~\ref{fig:1}). The reactions between O and O$^{18}$O (or O$^{17}$O) especially show strong deviations from mass-dependent behavior, resulting in high $\Delta^{17}$O values in O$_3$ in the upper atmosphere. In contrast, O$_3$ is relatively depleted in $^{18}$O and $^{17}$O in the lower atmosphere, due to the lower contribution of the O-O$^{18}$O (O$^{17}$O) reactions and the effect of CO$_2$ photolysis producing $^{18}$O- and $^{17}$O-depleted photochemical products. The slight decrease in $\delta ^{18}$O and $\delta ^{17}$O of O$_3$ with increasing altitude below $\sim 60$ km is caused by the temperature dependence of the fractionation associated with O$_3$ formation reaction (Figure~\ref{fig:1}).
	
	Atomic oxygen (O) is highly depleted in $^{18}$O and $^{17}$O, and its vertical isotopic profile is linked to that of O$_3$ (Figure~\ref{fig:2}). This relationship reflects the preferential incorporation of heavy oxygen isotopes into ozone during its formation, leaving atomic oxygen isotopically depleted. The depletion of $^{18}$O and $^{17}$O in atomic oxygen relative to molecular oxygen is also attributed to isotope exchange reactions. This depletion of the heavy oxygen isotopes in O directly affects the isotopic fractionation associated with atmospheric escape to space, which will be discussed in Section 4.3. The isotopic compositions of other species, including H$_2$O$_2$, OH, O$_2$, and HO$_2$, reflect the combined influence of both CO$_2$ photolysis, which preferentially depletes $^{18}$O and $^{17}$O, and ozone formation, which preferentially enriches these heavy isotopes.

\section{Discussion} \label{sec:discussion}
\begin{figure*}[htbp]
	\centering
	\includegraphics[width=2\columnwidth]{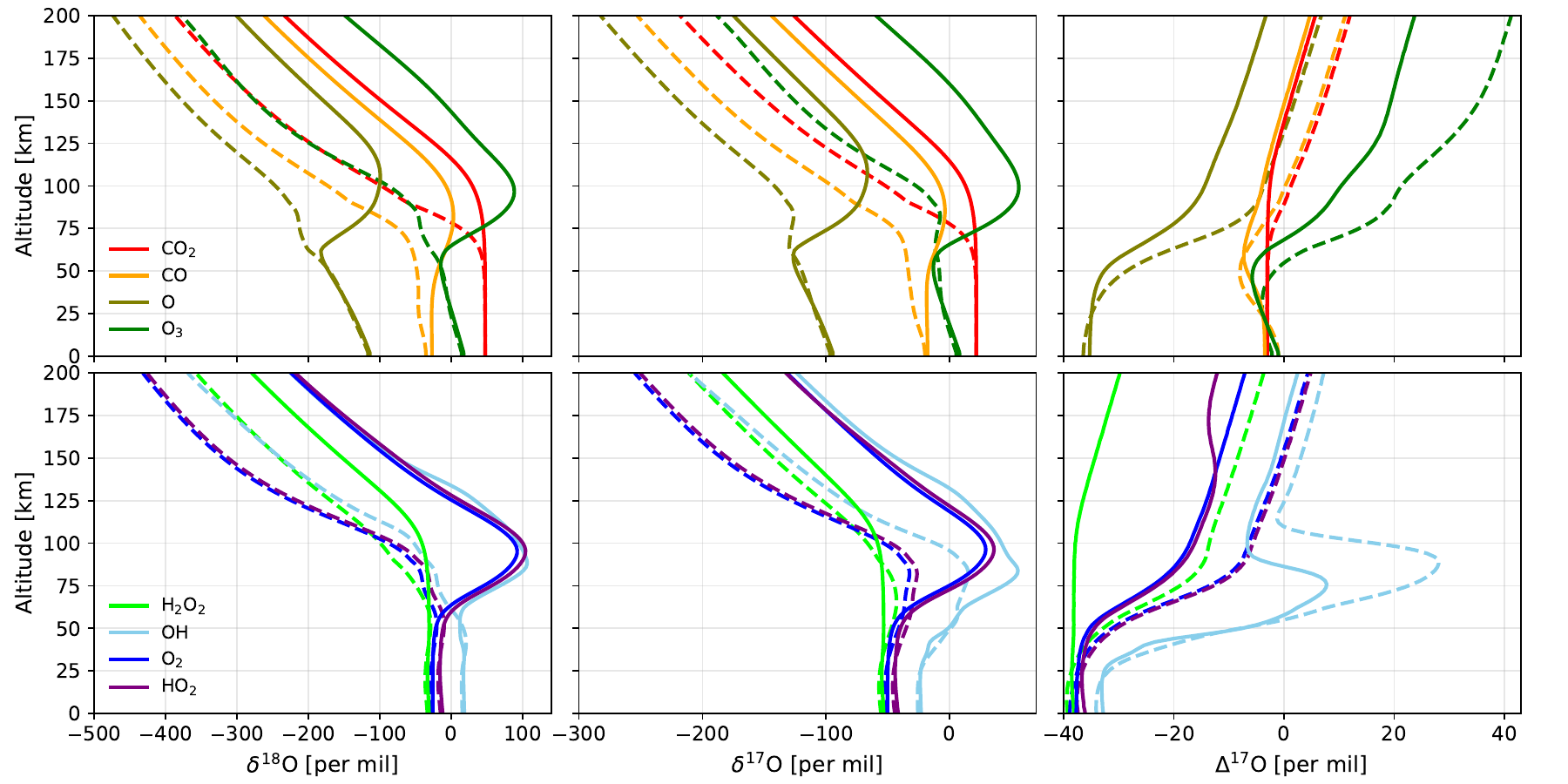}
	\caption{Vertical profiles of $\delta ^{18}$O (left), $\delta ^{17}$O, and $\Delta ^{17}$O, comparing the results obtained with the standard eddy diffusion coefficient (solid) and with an eddy diffusion coefficient one order of magnitude smaller (dashed). The upper panels show the results of CO$_2$, CO, O, and O$_{3}$, while the lower panels show those for H$_2$O$_2$, OH, O$_2$, and HO$_2$.}
	\label{fig:4}
\end{figure*}

\begin{figure*}[htbp]
	\centering
	\includegraphics[width=2\columnwidth]{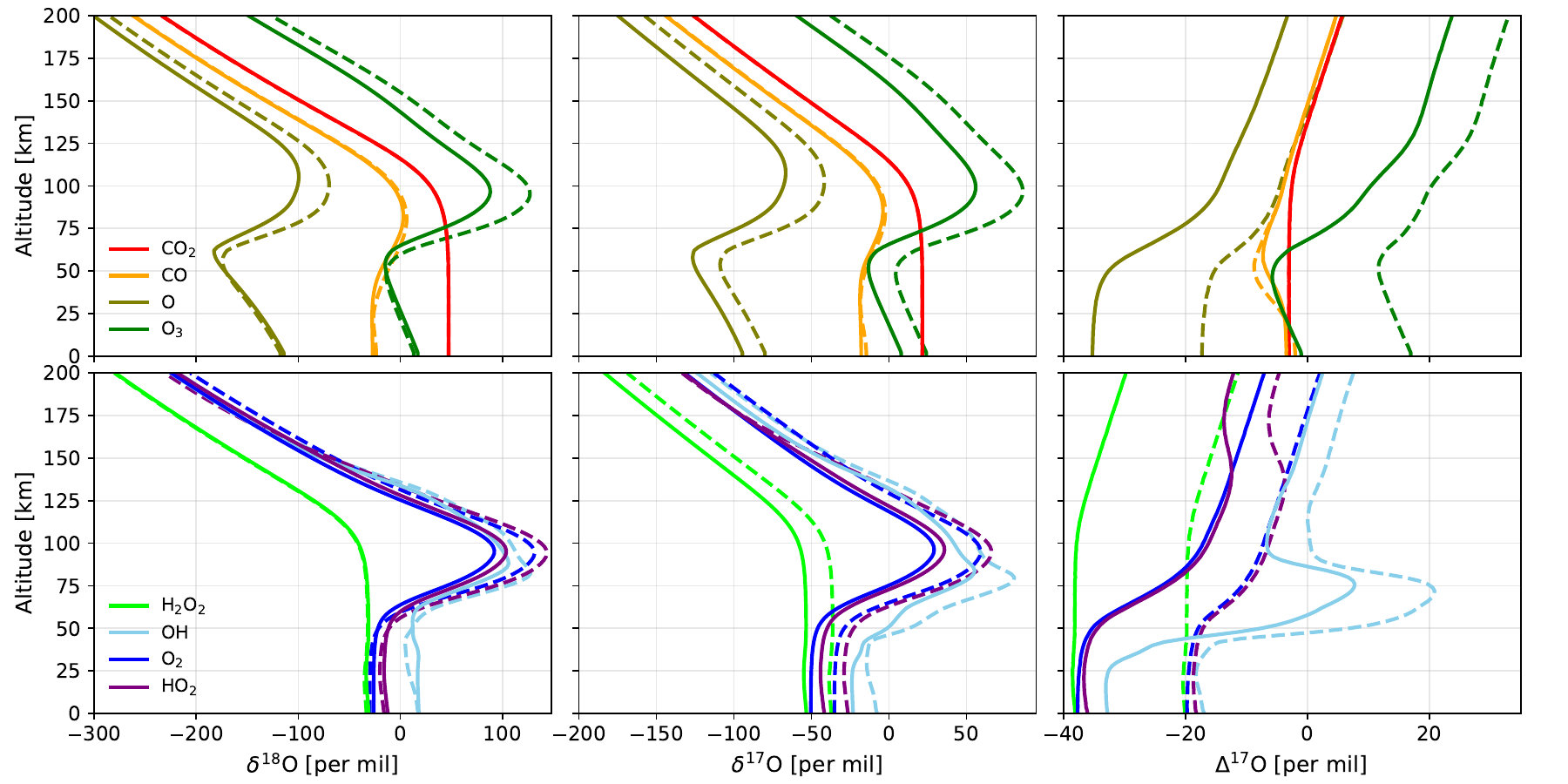}
	\caption{Vertical profiles of $\delta ^{18}$O (left), $\delta ^{17}$O (center), and $\Delta ^{17}$O (right), comparing the results obtained with the standard H$_2$O profile (solid), and with an H$_2$O mixing ratio 10 times larger (dashed). The upper panels show the results of CO$_2$, CO, O, and O$_{3}$, while the lower panels show those for H$_2$O$_2$, OH, O$_2$, and HO$_2$.}
	\label{fig:5}
\end{figure*}

\begin{figure*}[htbp]
	\centering
	\includegraphics[width=2\columnwidth]{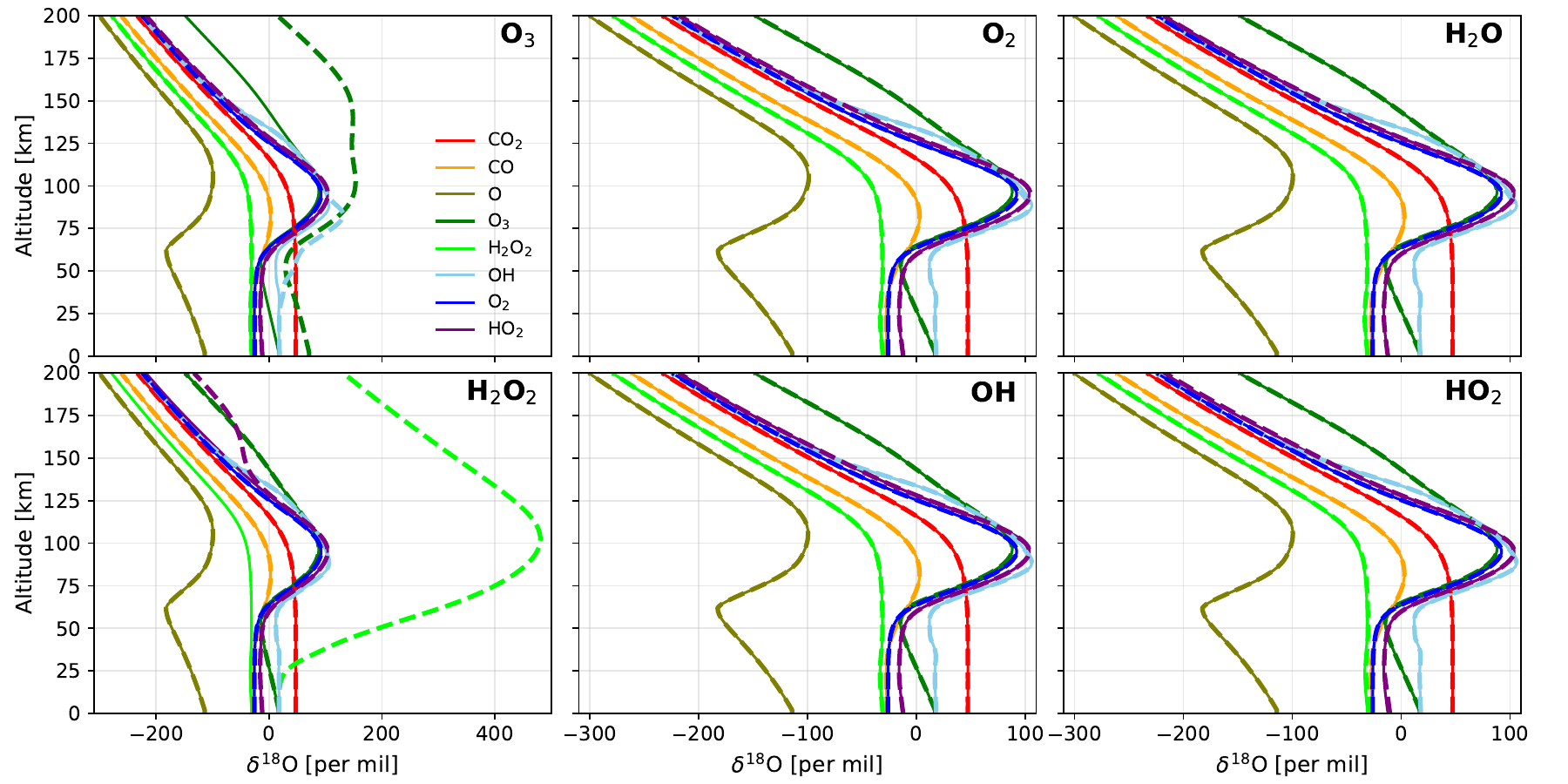}
	\caption{Vertical profiles of $\delta ^{18}$O, comparing the results obtained with the standard case (solid), and with artificial isotopic fractionation during the photolysis of O$_3$, O$_2$, H$_2$O, H$_2$O$_2$, OH, and HO$_2$ (dashed). The absorption cross sections of $^{18}$O-bearing isotopologues are reduced by 50‰ relative to those of the major isotopologues.}
	\label{fig:6}
\end{figure*}

\begin{figure*}[htbp]
	\centering
	\includegraphics[width=2\columnwidth]{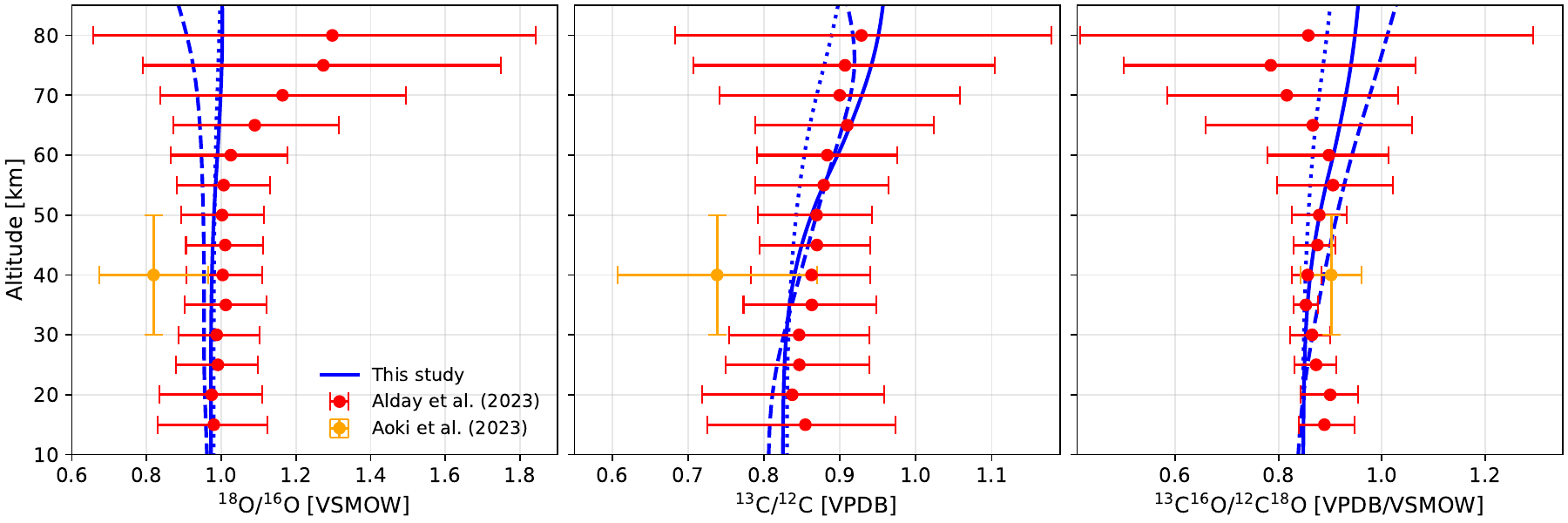}
	\caption{Comparison of measured isotopic profiles of CO from TGO solar occultation observations with model results. The left, center, and right panels show vertical profiles of $^{18}$O/$^{16}$O in CO, $^{13}$C/$^{12}$C in CO, and $^{13}$C$^{16}$O/$^{12}$C$^{18}$O, respectively. The isotopic ratios are normalized to the standard reference values of Vienna Standard Mean Ocean Water (VSMOW; $^{18}$O/$^{16}$O = $2.0052 \times 10^{-3}$) and Vienna Pee Dee Belemnite (VPDB; $^{13}$C/$^{12}$C = $1.123 \times 10^{-2}$). The blue lines represent the calculation results: the solid, dashed, and dotted lines correspond to the results with the standard eddy diffusion coefficient, a coefficient one order of magnitude smaller, and a coefficient one order of magnitude larger, respectively. The red and orange bars represent the values retrieved from TGO solar occultation measurements by \citet{Alday2023} and \citet{Aoki2023}, respectively.}
	\label{fig:7}
\end{figure*}

\subsection{Parameter dependence}	
\subsubsection{Eddy diffusion coefficient}
	The magnitude of the eddy diffusion coefficient varies substantially with season and latitude \citep[e.g.,][]{Slipski2018,Yoshida2022}, which can influence the atmospheric profile including the oxygen isotopic compositions. Figure~\ref{fig:4} compares the isotopic profiles obtained using an eddy diffusion coefficient that is one order of magnitude smaller than the standard value. The corresponding profile of the eddy diffusion coefficient is shown in Figure A2. Under the suppressed eddy diffusion condition, diffusive separation between isotopologues through molecular diffusion becomes more effective relative to the homogenization via eddy diffusion. As a result, both $\delta ^{18}$O and $\delta ^{17}$O for all the O-bearing species decrease in the upper region. Conversely, under a higher eddy diffusion coefficient, vertical mixing more effectively homogenizes the isotopic composition, thereby suppressing molecular diffusive separation. These changes in the isotopic compositions at high altitudes can modify the fractionation associated with atmospheric escape, as discussed in Section 4.3.

\subsubsection{H$_2$O mixing ratio}
	In this study, the vertical profile of H$_2$O was prescribed so as to reproduce the globally averaged water vapor abundance (Section 2). In reality, however, both the total water vapor abundance and its vertical distribution vary substantially with latitude and season \citep[e.g.,][]{Smith2002,Fedorova2006,Crismani2021,Aoki2022}. Variations in the H$_2$O abundance in the Martian atmosphere can also affect the oxygen isotopic composition. Figure~\ref{fig:5} shows the oxygen isotopic composition profiles obtained using an H$_2$O mixing ratio 10 times larger than the standard value. The H$_2$O number density profile is shown in Figure A3. Under the H$_2$O-rich condition, both $\delta^{18}$O and $\delta^{17}$O of photochemical products increase, because H$_2$O, which has higher $\delta^{18}$O and $\delta^{17}$O values than CO$_2$ (Section 2), produces HO$_x$ radicals enriched in $^{18}$O and $^{17}$O through photolysis. Although the abundance of O$_3$ decreases due to its anti-correlation with H$_2$O \citep[e.g.,][]{Lefevre2004,Perrier2006,Olsen2022}, this does not significantly diminish the impact of O$_3$-related isotopic fractionation on the isotopic compositions of other O-bearing species, such as atomic oxygen. This is because an increase in H$_2$O also lowers the abundances of other O-bearing species, such that the relative importance of O$_3$-related fractionation processes remains significant. The overall influence of changes in H$_2$O abundance on the oxygen isotopic composition profiles is modest compared to the effect of changes in the eddy diffusion coefficient. This is because the contribution of H$_2$O as a source of O-bearing photochemical products remains limited, even when its abundance is increased.

\subsubsection{Fractionation via photolysis of species other than CO$_2$}
	We have not explicitly considered isotopic fractionation associated with photolysis of O-bearing species other than CO$_2$, primarily due to the limited availability and large uncertainties of cross-section data of their minor isotopes. Nevertheless, photolysis of species such as O$_3$ and O$_2$ could, in principle, influence the oxygen isotopic composition \citep[e.g.,][]{Yoshino1987,Liang2004,Miller2005,Liang2006,Ndengue2014,Huang2019,Danielache2023}. To evaluate the potential impact of such processes, we perform sensitivity tests by artificially imposing isotopic fractionation during the photolysis of minor species (Figure~\ref{fig:6}). Specifically, the absorption cross sections of $^{18}$O-bearing isotopologues are reduced by 50‰ relative to those of the major isotopologues. This magnitude is comparable to typical values predicted by zero-point-energy (ZPE)-based models \citep{Miller2000,Miller2005,Liang2004}. Overall, the effects on the atmospheric oxygen isotopic compositions are small. In the cases where isotopic fractionation is introduced for the photolysis of O$_2$, H$_2$O, OH, and HO$_2$, the resulting changes in the isotopic compositions of the O-bearing species are negligible, because photolysis of these species constitutes only a minor contribution to the overall oxygen budget. As an exception, the isotopic compositions of O$_3$ and H$_2$O$_2$ are more sensitive to their own photolysis-induced fractionation, particularly in upper regions. Both species are primarily produced at lower altitudes and subsequently transported upward, where they are efficiently decomposed by photolysis. As a result of the preferential removal of lighter isotopologues during photolysis, heavier isotopes become relatively enriched in the remaining molecules at high altitudes.	

\subsection{Comparison with the TGO observation results}
	\citet{Alday2023} and \citet{Aoki2023} retrieved the vertical profiles of the carbon and oxygen isotopic compositions of CO from solar occultation measurements conducted by ExoMars Trace Gas Orbiter (TGO). We compare our calculation results of the CO isotopic compositions with their observational results. Here, we also use the calculation results of carbon isotopic composition profiles reported by \citet{Yoshida2023}. The overall trend of the vertical profiles of both oxygen and carbon isotopic composition, where the isotopic fractionation of carbon is entirely more pronounced by that of oxygen, is in good agreement between our model and the TGO measurements (Figure~\ref{fig:7}). The vertical profile of the $^{13}$C$^{16}$O/$^{12}$C$^{18}$O ratio, which can be retrieved with relatively weak temperature dependence and smaller uncertainties \citep{Aoki2023}, is also well reproduced by our model. This agreement supports the interpretation that the isotopic composition of CO is primarily controlled by isotopic fractionation associated with CO$_2$ photolysis.
	
	Although the $^{13}$C/$^{12}$C and $^{18}$O/$^{16}$O ratios retrieved by \citet{Aoki2023} are slightly lower than the modeled values, they overlap with our calculation results within their relatively large standard deviation ranges. We note that the observed $^{13}$C/$^{12}$C and $^{18}$O/$^{16}$O ratios have relatively large uncertainty due to strong temperature dependence in the $^{12}$C$^{16}$O retrievals, as discussed in detail by \citet{Aoki2023}. Therefore, for comparison with our results, we use the altitude-averaged isotopic ratios reported in Table 2 of \citet{Aoki2023}, rather than the individual altitude profiles. At higher altitudes above $\sim 60$ km, the model does not fully reproduce the increase in the $^{18}$O/$^{16}$O ratio suggested by the TGO observations, while the modeled values remain within the observational uncertainty range. The observations could be affected by the lower SNR on the $^{12}$C$^{18}$O retrievals represented by the error bars. Further observational constraints are required to clarify this potential discrepancy. In addition, since this model calculates a steady-state vertical profile at a solar zenith angle of zero and neglects temporal and spatial variability, these simplifications may represent one possible source of the remaining discrepancies with the TGO observations. 
	
	We also compare our treatment of CO$_2$ isotopic compositions with previous TGO measurements. \citet{Alday2021} retrieved vertical profiles of CO$_2$ isotopic ratios from TGO observations between 70 and 130 km. They showed that the $^{13}$C/$^{12}$C and $^{18}$O/$^{16}$O ratios decrease above $\sim$100 km. This trend is consistent with diffusive separation above the homopause, as also suggested by our results (Figure~\ref{fig:2}; see also Figure 2 of \citet{Yoshida2023} for the carbon isotopic composition). \citet{Alday2021} also noted that the average CO$_2$ isotopic ratios retrieved at 70--90 km do not show clear enrichment in heavy isotopes relative to Earth-like values, with $\delta^{18}$O $= -29 \pm 38$‰, $\delta^{17}$O $= -11 \pm 41$‰, and $\delta^{13}$C $= -3 \pm 37$‰. In contrast, the calculated CO$_2$ isotopic ratios in our model remain close to the prescribed surface values enriched in heavy isotopes (Figure~\ref{fig:2}), which are assumed to be equal to those measured by SAM/TLS on the Curiosity Rover, as described in Section 2. The origin of this difference between the lower and upper atmosphere remains unclear. \citet{Alday2021} discussed several possible explanations. For example, climatological and surface--atmosphere fractionation effects may influence the representative isotopic ratios derived from each data set. Because SAM/TLS measurements were obtained at a fixed location and over limited ranges of season and local time, they may not represent the global atmospheric reservoir, whereas TGO observations sample wider ranges of latitudes, seasons, and local times.

\begin{figure}[htbp]
	\centering
	\includegraphics[width=\columnwidth]{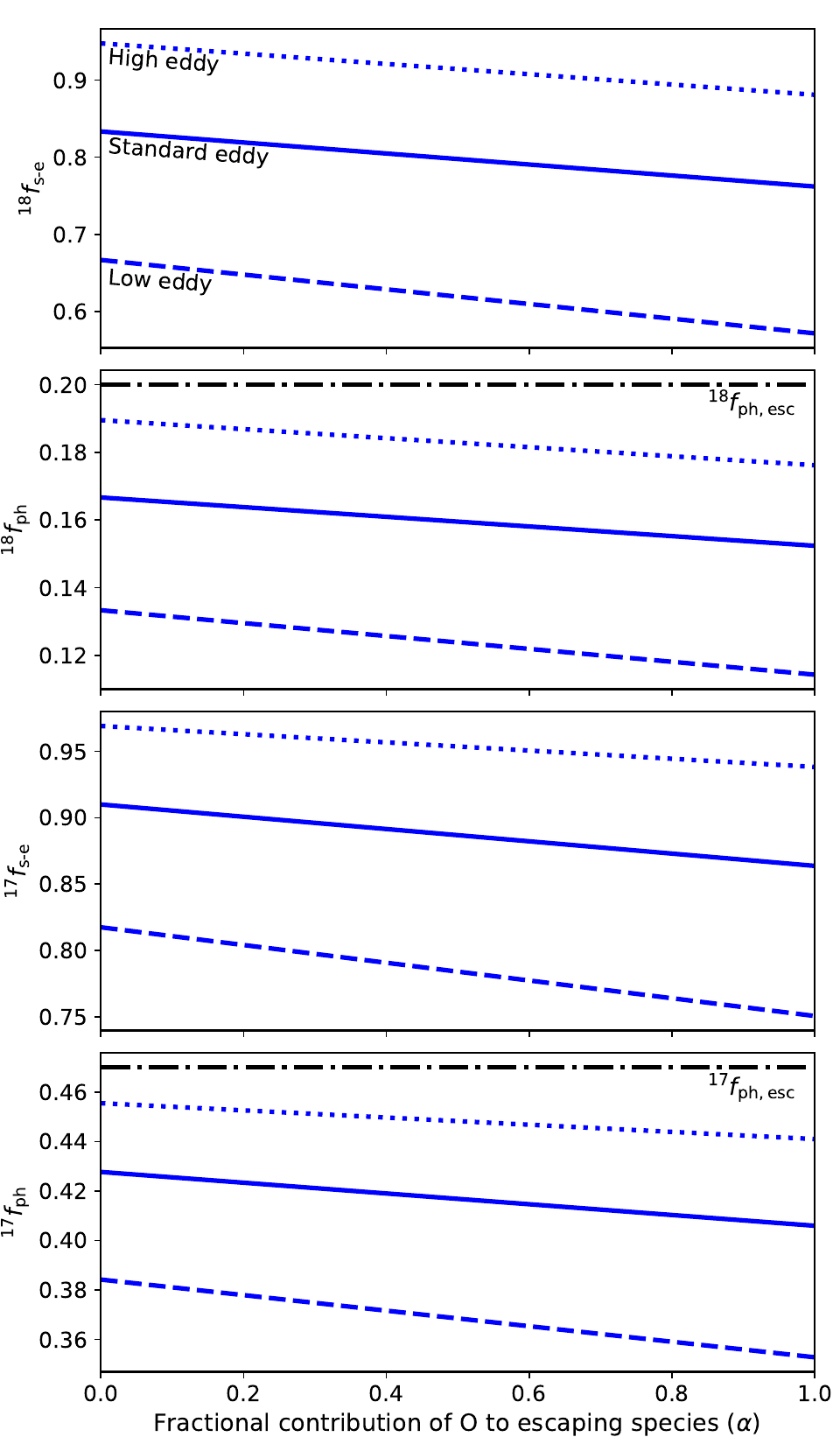}
	\caption{Fractionation factor associated with oxygen escape. From top to bottom, the panels show $^{18}f_{\mathrm{s\text{-}e}}$, $^{18}f_{\rm ph}$, $^{17}f_{\mathrm{s\text{-}e}}$, and $^{17}f_{\rm ph}$. The definitions of these fractionation factors are described in Section 4.3. The horizontal axis is the fractional contribution of atomic oxygen to O-bearing escaping species, corresponding to $\alpha$ in Equation (4). The solid, dashed, and dotted blue lines represent the results obtained using the standard eddy diffusion coefficient, a coefficient one order of magnitude smaller, and a coefficient one order of magnitude larger, respectively. The black dash-dotted horizontal lines in the second and fourth panels represent $^{18}f_{\rm ph,esc}$ and $^{17}f_{\rm ph,esc}$, respectively.}
	\label{fig:8}
\end{figure}

\subsection{Oxygen isotopic fractionation via atmospheric escape}
	In this section, we estimate the fractionation factor for oxygen escape while accounting for the chemical isotopic fractionation. As shown in Figure~\ref{fig:2}, both O and CO$_2$, which are sources of escaping species, are depleted in $^{18}$O and $^{17}$O in the upper region where atmospheric escape occurs. This depletion can suppress the escape of $^{18}$O and $^{17}$O and thereby enhances the net oxygen isotopic fractionation.
	
	The major process of oxygen escape from the present-day Martian atmosphere is estimated to be photochemical escape via dissociative recombination of O$_{2}^{+}$ \citep[e.g.,][]{Jakosky2018}:
	\begin{equation}
		\mathrm{O_{2}^{+}}+\mathrm{e} \to \mathrm{O}+\mathrm{O}.
	\end{equation}
	The fractionation factor for photochemical escape can be expressed as
	\begin{equation}
		^{j}f_{\rm ph}=\, ^{j}f_{\mathrm{s\text{-}e}} \times \, ^{j}f_{\mathrm{ph,esc}},
	\end{equation}
	\begin{equation}
		^{j}f_{\mathrm{s\text{-}e}} = \frac{^{j}R_{\rm esc}}{^{j}R_{\rm surf}},
	\end{equation}
	where $j(=17,18)$ is the mass number of the minor oxygen isotopes, $^{j}f_{\rm ph}$ is the net fractionation factor of $^{j}$O via photochemical escape, $^{j}f_{\rm s\text{-}e}$ is the fractionation factor of $^{j}$O between the surface and escape region, $^{j}f_{\rm ph,esc}$ is the fractionation factor of $^{j}$O associated with the photochemical escape mechanism, $^{j}R_{\rm esc}$ is the $^{j}$O/$^{16}$O ratio of the source of escaping species, and $^{j}R_{\rm surf}$ is the $^{j}$O/$^{16}$O ratio of CO$_2$ at the surface, which is the representative value of the bulk atmospheric resorvoir. Because our model does not fully treat ion chemistry, which would be required to directly compute the isotopic composition of escaping species self-consistently across multiple reaction pathways, we introduce a parameter $\alpha$ as an effective factor representing the fractional contribution of atomic O to escaping oxygen species. The isotopic ratio of escaping oxygen is then approximated as
	\begin{equation}
		^{j}R_{\rm esc}=  \alpha\, ^{j}R_{\rm O} + (1-\alpha) \, ^{j}R_{\rm CO_{2}},
	\end{equation}
	where $^{j}R_{i}$ is the $^{j}$O/$^{16}$O ratio of species $i$. We also assume that the altitude of the escape region is 160 km, where the production of escaping oxygen atoms typically peaks, and $^{18}f_{\rm ph,esc}=0.2$ \citep[][]{Fox2009,Fox2010}. Since $^{17}f_{\rm ph,esc}$ has not been estimated previously, we suppose $^{17}f_{\rm ph,esc}=0.47$ based on mass-dependent scaling. As shown in Figure~\ref{fig:8}, both $^{18}f_{\rm ph}$ and $^{17}f_{\rm ph}$ are further reduced by the depletion of $^{18}$O and $^{17}$O in the escape region through the chemical isotopic fractionation, along with the effect of diffusive separation shown by \citet{Brinjikji2021,Lyons2024}. This indicates that escaping oxygen via photochemical escape is highly isotopically fractionated. $^{j}f_{\rm s\text{-}e}$ and $^{j}f_{\rm ph}$ slightly decrease with increasing $\alpha$, because O is more depleted in the heavier oxygen isotopes than CO$_2$ (Figure~\ref{fig:2}). Enhanced diffusive separation under reduced eddy diffusion also increases the degree of fractionation.
	
	In addition to photochemical escape, oxygen is lost from Mars through ion escape driven by solar wind interaction, and through sputtering caused by the impact of energetic precipitating ions \citep[e.g.,][]{Brain2015,Jakosky2018}. Supposing that the mass fractionation by these processes themselves is negligible \citep{Chassefiere2004}, the isotopic fractionation depends solely on the difference between the isotopic composition at the surface and that in the escape region. Thus, the fractionation factor can be approximated by $^{j}f_{\rm esc}$ (Figure~\ref{fig:8}), which deviates significantly from the photochemical escape fractionation factor, $^{j}f_{\rm ph}$. This contrast suggests a potential dichotomy in the isotopic composition of escaping oxygen, depending on the escape mechanism.
	
	Isotopic compositions of escaping oxygen may be validated by the Martian Moons eXploration (MMX) mission, which is planned by the Japan Aerospace Exploration Agency (JAXA) to target the two Martian moons \citep{Kuramoto2022}. The mass spectrum analyzer on board MMX, with unprecedented mass resolution, will be able to measure the isotope ratios of escaping ion species such as O$^{+}$ and C$^{+}$ \citep{Yokota2021,Kuramoto2022,Ogohara2022,Yokota2025}. Such observations will provide crucial constraints on the isotopic fractionation associated with atmospheric escape, and thus improve our understanding of the long-term evolution of the Martian atmosphere.
	
\section{Conclusions}
	We have developed a 1D photochemical model that incorporates oxygen isotopic fractionation associated with CO$_2$ photolysis and O$_3$ formation, and applied it to investigate the vertical profiles of oxygen isotopic compositions in the Martian atmosphere. Our calculations show that CO is depleted in heavy oxygen isotopes relative to CO$_2$, reaching $\delta ^{18}$O $\sim -25$‰ and $\delta ^{17}$O $\sim -15$‰, primarily due to isotopic fractionation during CO$_2$ photolysis. The vertical profiles of oxygen and carbon isotopic compositions are in good agreement between our model and the TGO measurements, supporting the interpretation that the isotopic composition of CO is primarily controlled by isotopic fractionation associated with CO$_2$ photolysis. O$_3$ is strongly enriched in $^{18}$O and $^{17}$O, reaching $\delta ^{18}$O $\sim 100$‰ and $\delta ^{17}$O $\sim 50$‰ as a consequence of the isotopic fractionation during its formation, whereas atomic oxygen is highly depleted in the heavy oxygen isotopes with $\delta ^{18}$O $\lesssim -100$‰ and $\delta ^{17}$O $\lesssim -50$‰ so as to compensate for their enrichment in O$_3$. These chemical fractionation processes lead to the depletion of $^{18}$O and $^{17}$O in the oxygen species present in the escape region, thereby enhancing the isotopic fractionation associated with oxygen escape to space. Such fractionated isotopic compositions of escaping oxygen may be detectable by the Martian Moons eXploration (MMX) mission.
	
\bibliographystyle{aasjournal} 
\bibliography{references} 	
	
\appendix
\section{Supplemental figures}
	The fixed profiles of temperature, eddy diffusion coefficient, and H$_2$O number density are shown in Figure A1, A2, and A3, respectively. The calculated number density profiles are shown in Figure A4. The calculated profiles of the oxygen isotopic compositions comparing the results obtained with solar zenith angle of 0 degrees and 60 degrees are shown in Figure A5.

\section{Chemical reactions}
	The chemical reactions used in our photochemical model are listed in Table A1.

\begin{figure}[htbp]
	\centering
	\includegraphics[width=0.4\columnwidth]{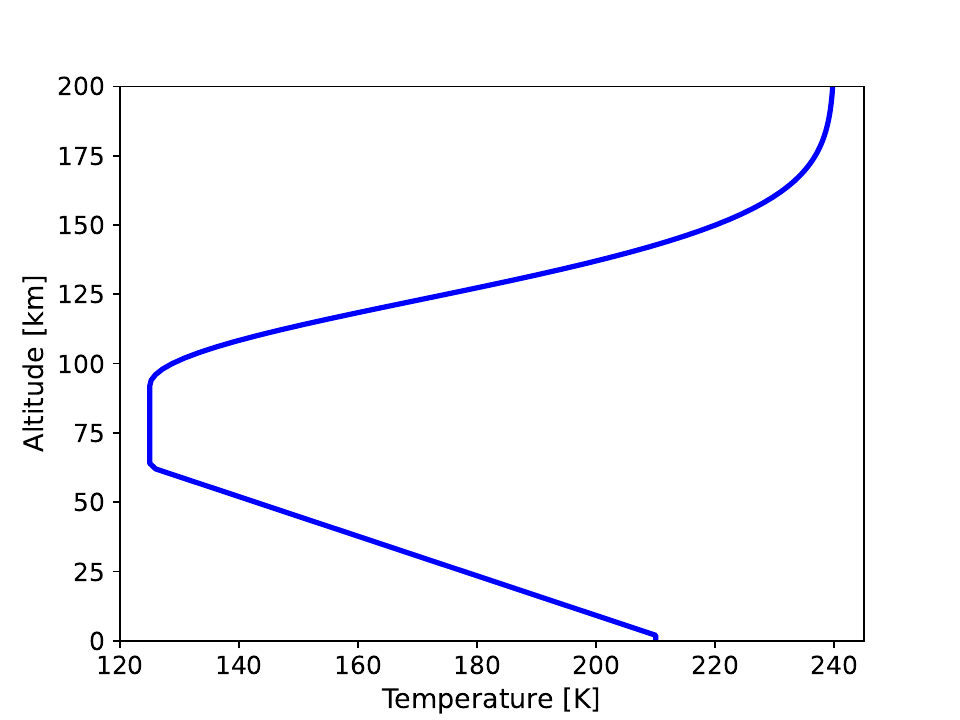} \\
	{\bfseries Figure A1.} Temperature profile.
	\label{fig:A1}
\end{figure}

\begin{figure}[htbp]
	\centering
	\includegraphics[width=0.4\columnwidth]{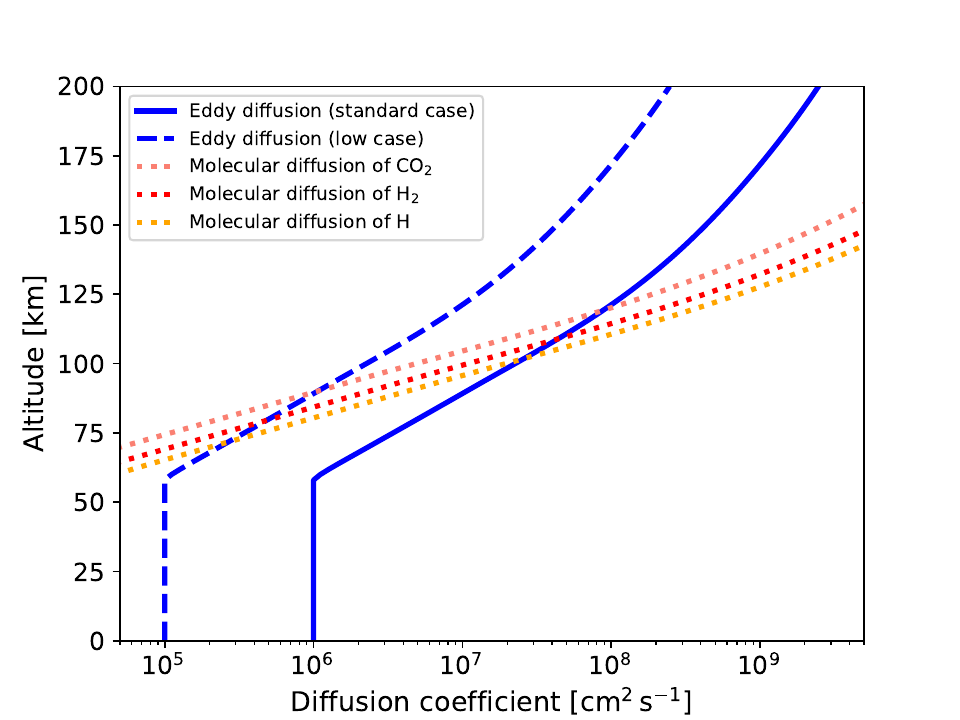} \\
	{\bfseries Figure A2.} Vertical profiles of the eddy diffusion coefficient and molecular diffusion coefficient. The solid blue line represents the profile with the standard eddy diffusion coefficient, while the dashed blue line represents the profile with an coefficient one order of magnitude smaller used in Section 4.1. The dotted orange, red, and pink lines represent the profiles of the molecular diffusion coefficients of CO$_2$, H$_2$, and H, respectively.
	\label{fig:A2}
\end{figure}

\begin{figure}[htbp]
	\centering
	\includegraphics[width=0.4\columnwidth]{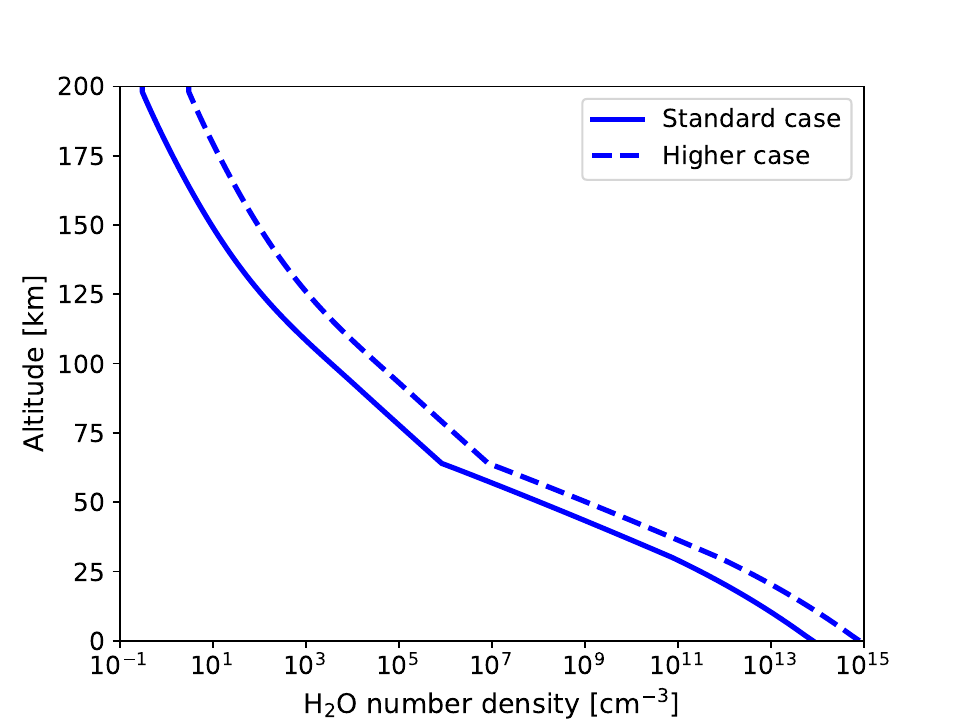} \\
	{\bfseries Figure A3.} H$_2$O number density profile. The solid blue line represents the stanrard profile, while the dashed blue line represents that 10 times as high as the standard setting used in Section 4.1. 
	\label{fig:A3}
\end{figure}

\begin{figure}[htbp]
	\centering
	\includegraphics[width=0.9\columnwidth]{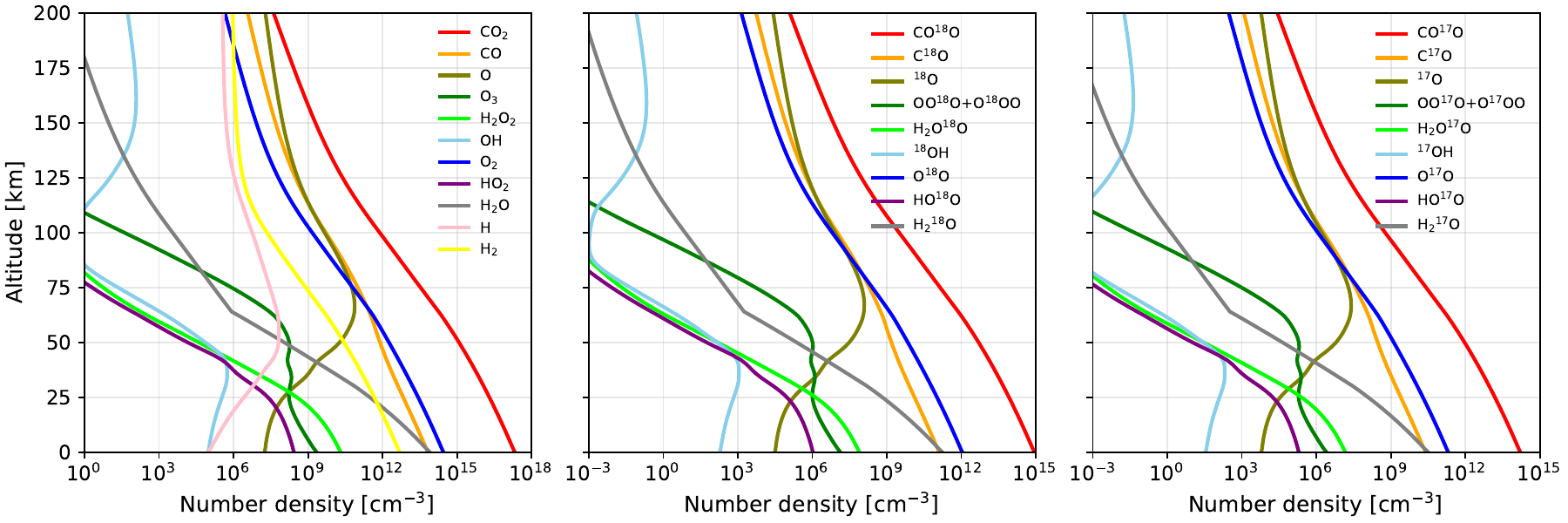} \\
	{\bfseries Figure A4.} The calculated number density profiles.
	\label{fig:A4}
\end{figure}

\begin{figure}[htbp]
	\centering
	\includegraphics[width=0.9\columnwidth]{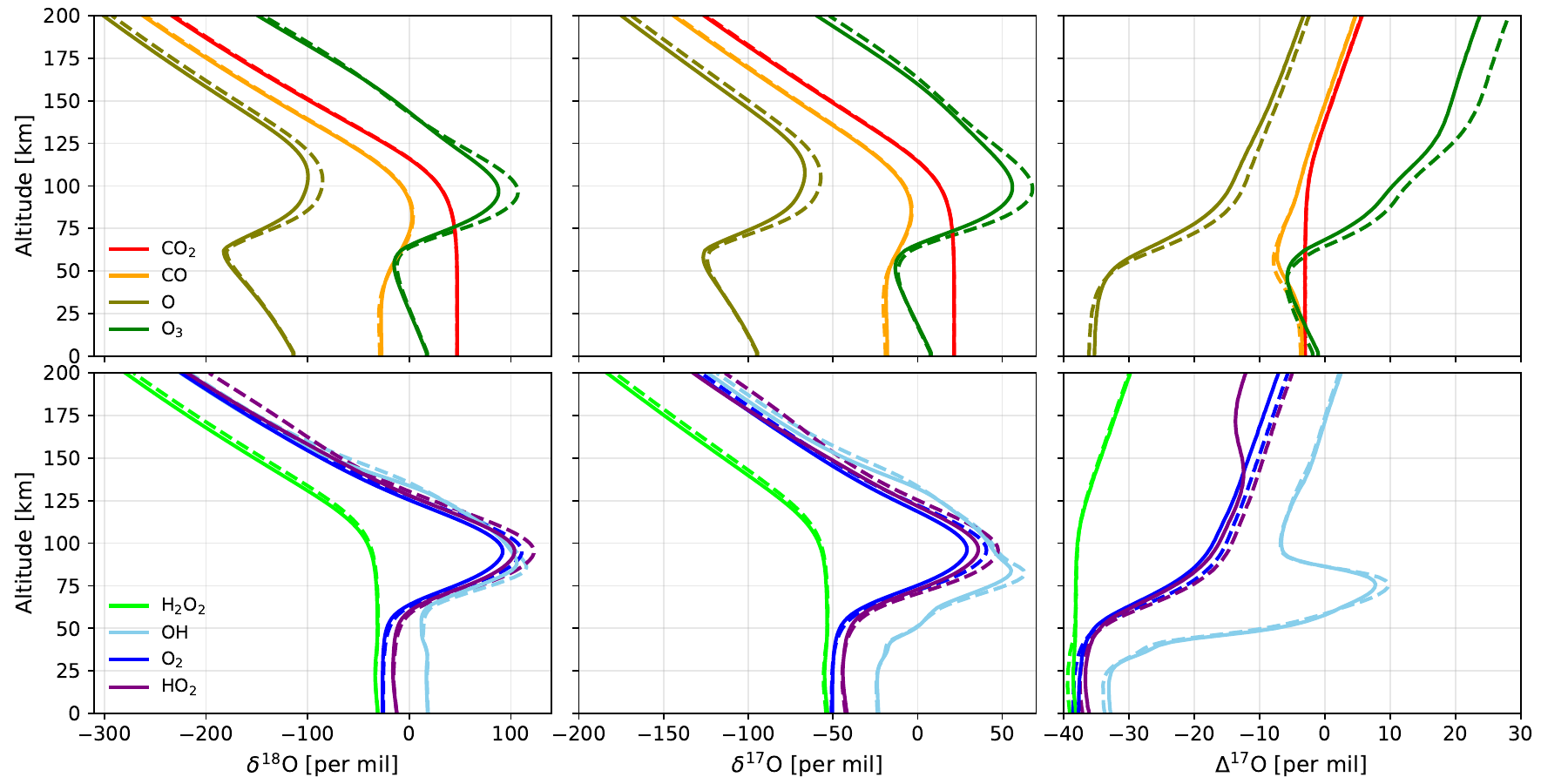} \\
	{\bfseries Figure A5.} Vertical profiles of $\delta ^{18}$O (left), $\delta ^{17}$O, and $\Delta ^{17}$O, comparing the results obtained with solar zenith angle of 0 degrees (solid) and 60 degrees (dashed). The upper panels show the results of CO$_2$, CO, O, and O$_{3}$, while the lower panels show those for H$_2$O$_2$, OH, O$_2$, and HO$_2$.
	\label{fig:A5}
\end{figure}

\clearpage
\setcounter{table}{0}
\renewcommand{\thetable}{A\arabic{table}}
\begin{longtable}{llllll}
\caption{Chemical reaction list.}\\
\hline
No. & Reaction & & & Rate coefficient (2-body: cm$^{3}$ s$^{-1}$; 3-body: cm$^{6}$ s$^{-1}$) & Reference \\
\hline
R1  & $\mathrm{CO}_{2}+h\nu$ & $\to$ & $\mathrm{CO}+\mathrm{O}$ & Photodissociation & a\\
R2  & $\mathrm{CO}_{2}+h\nu$ & $\to$ & $\mathrm{CO}+\mathrm{O}(^1\mathrm{D})$ & Photodissociation & a\\
R3  & $\mathrm{H}_{2}\mathrm{O}+h\nu$ & $\to$ & $\mathrm{H}+\mathrm{OH}$ & Photodissociation & a\\
R4  & $\mathrm{H}_{2}\mathrm{O}+h\nu$ & $\to$ & $\mathrm{H}_{2}+\mathrm{O}(^1\mathrm{D})$ & Photodissociation & a\\
R5  & $\mathrm{O}_{3}+h\nu$ & $\to$ & $\mathrm{O}_{2}+\mathrm{O}$ & Photodissociation & a\\
R6  & $\mathrm{O}_{3}+h\nu$ & $\to$ & $\mathrm{O}_{2}+\mathrm{O}(^1\mathrm{D})$ & Photodissociation & a\\
R7  & $\mathrm{O}_{2}+h\nu$ & $\to$ & $\mathrm{O}+\mathrm{O}$ & Photodissociation & a\\
R8  & $\mathrm{O}_{2}+h\nu$ & $\to$ & $\mathrm{O}+\mathrm{O}(^1\mathrm{D})$ & Photodissociation & a\\
R9  & $\mathrm{H}_{2}+h\nu$ & $\to$ & $\mathrm{H}+\mathrm{H}$ & Photodissociation & a\\
R10 & $\mathrm{OH}+h\nu$ & $\to$ & $\mathrm{O}+\mathrm{H}$ & Photodissociation & a\\
R11 & $\mathrm{OH}+h\nu$ & $\to$ & $\mathrm{O}(^1\mathrm{D})+\mathrm{H}$ & Photodissociation & a\\
R12 & $\mathrm{HO}_{2}+h\nu$ & $\to$ & $\mathrm{OH}+\mathrm{O}$ & Photodissociation & a\\
R13 & $\mathrm{H}_{2}\mathrm{O}_{2}+h\nu$ & $\to$ & $\mathrm{OH}+\mathrm{OH}$ & Photodissociation & a\\
R14 & $\mathrm{H}_{2}\mathrm{O}_{2}+h\nu$ & $\to$ & $\mathrm{HO}_{2}+\mathrm{H}$ & Photodissociation & a\\
R15 & $\mathrm{H}_{2}\mathrm{O}_{2}+h\nu$ & $\to$ & $\mathrm{H}_{2}\mathrm{O}+\mathrm{O}(^1\mathrm{D})$ & Photodissociation & a\\
R16 & $\mathrm{O}+\mathrm{O}+\mathrm{M}$ & $\to$ & $\mathrm{O}_{2}+\mathrm{M}$ & $5.4\times10^{-33}(300/T)^{3.25}$ & a\\
R17 & $\mathrm{O}+\mathrm{O}_{2}+\mathrm{CO}_{2}$ & $\to$ & $\mathrm{O}_{3}+\mathrm{CO}_{2}$ & $1.5\times10^{-33}(300/T)^{2.4}$ & a\\
R18 & $\mathrm{O}+\mathrm{O}_{3}$ & $\to$ & $\mathrm{O}_{2}+\mathrm{O}_{2}$ & $8.0\times10^{-12}\exp(-2060/T)$ & a\\
R19 & $\mathrm{O}+\mathrm{CO}+\mathrm{M}$ & $\to$ & $\mathrm{CO}_{2}+\mathrm{M}$ & $2.2\times10^{-33}\exp(-1780/T)$ & a\\
R20 & $\mathrm{O}(^1\mathrm{D})+\mathrm{O}_{2}$ & $\to$ & $\mathrm{O}+\mathrm{O}_{2}$ & $3.2\times10^{-11}\exp(70/T)$ & a\\
R21 & $\mathrm{O}(^1\mathrm{D})+\mathrm{O}_{3}$ & $\to$ & $\mathrm{O}_{2}+\mathrm{O}_{2}$ & $1.2\times10^{-10}$ & a\\
R22 & $\mathrm{O}(^1\mathrm{D})+\mathrm{O}_{3}$ & $\to$ & $\mathrm{O}+\mathrm{O}+\mathrm{O}_{2}$ & $1.2\times10^{-10}$ & a\\
R23 & $\mathrm{O}(^1\mathrm{D})+\mathrm{H}_{2}$ & $\to$ & $\mathrm{H}+\mathrm{OH}$ & $1.2\times10^{-10}$ & a\\
R24 & $\mathrm{O}(^1\mathrm{D})+\mathrm{CO}_{2}$ & $\to$ & $\mathrm{O}+\mathrm{CO}_{2}$ & $7.5\times10^{-11}\exp(115/T)$ & a\\
R25 & $\mathrm{O}(^1\mathrm{D})+\mathrm{H}_{2}\mathrm{O}$ & $\to$ & $\mathrm{OH}+\mathrm{OH}$ & $1.63\times10^{-10}\exp(60/T)$ & a\\
R26 & $\mathrm{H}_{2}+\mathrm{O}$ & $\to$ & $\mathrm{OH}+\mathrm{H}$ & $6.34\times10^{-12}\exp(-4000/T)$ & a\\
R27 & $\mathrm{OH}+\mathrm{H}_{2}$ & $\to$ & $\mathrm{H}_{2}\mathrm{O}+\mathrm{H}$ & $9.01\times10^{-13}\exp(-1526/T)$ & a\\
R28 & $\mathrm{H}+\mathrm{H}+\mathrm{CO}_{2}$ & $\to$ & $\mathrm{H}_{2}+\mathrm{CO}_{2}$ & $1.6\times10^{-32}(298/T)^{2.27}$ & a\\
R29 & $\mathrm{H}+\mathrm{OH}+\mathrm{CO}_{2}$ & $\to$ & $\mathrm{H}_{2}\mathrm{O}+\mathrm{CO}_{2}$ & $1.292\times10^{-30}(300/T)^{2}$ & a\\
R30 & $\mathrm{H}+\mathrm{HO}_{2}$ & $\to$ & $\mathrm{OH}+\mathrm{OH}$ & $7.2\times10^{-11}$ & a\\
R31 & $\mathrm{H}+\mathrm{HO}_{2}$ & $\to$ & $\mathrm{H}_{2}\mathrm{O}+\mathrm{O}(^1\mathrm{D})$ & $1.6\times10^{-12}$ & a\\
R32 & $\mathrm{H}+\mathrm{HO}_{2}$ & $\to$ & $\mathrm{H}_{2}+\mathrm{O}_{2}$ & $3.45\times10^{-12}$ & a\\
R33 & $\mathrm{H}+\mathrm{H}_{2}\mathrm{O}_{2}$ & $\to$ & $\mathrm{HO}_{2}+\mathrm{H}_{2}$ & $2.8\times10^{-12}\exp(-1890/T)$ & a\\
R34 & $\mathrm{H}+\mathrm{H}_{2}\mathrm{O}_{2}$ & $\to$ & $\mathrm{H}_{2}\mathrm{O}+\mathrm{OH}$ & $1.7\times10^{-11}\exp(-1800/T)$ & a\\
R35 & $\mathrm{H}+\mathrm{O}_{2}+\mathrm{M}$ & $\to$ & $\mathrm{HO}_{2}+\mathrm{M}$ & $k_{0}=8.8\times10^{-32}(300/T)^{1.3}$, $k_{\infty}=7.5\times 10^{-11} (300/T)^{-0.2}$ & a\\
R36 & $\mathrm{H}+\mathrm{O}_{3}$ & $\to$ & $\mathrm{OH}+\mathrm{O}_{2}$ & $1.4\times10^{-10}\exp(-470/T)$ & a\\
R37 & $\mathrm{O}+\mathrm{OH}$ & $\to$ & $\mathrm{O}_{2}+\mathrm{H}$ & $1.8\times10^{-11}\exp(180/T)$ & a\\
R38 & $\mathrm{O}+\mathrm{HO}_{2}$ & $\to$ & $\mathrm{OH}+\mathrm{O}_{2}$ & $3.0\times10^{-11}\exp(200/T)$ & a\\
R39 & $\mathrm{O}+\mathrm{H}_{2}\mathrm{O}_{2}$ & $\to$ & $\mathrm{OH}+\mathrm{HO}_{2}$ & $1.4\times10^{-12}\exp(-2000/T)$ & a\\
R40 & $\mathrm{OH}+\mathrm{OH}$ & $\to$ & $\mathrm{H}_{2}\mathrm{O}+\mathrm{O}$ & $1.8\times10^{-12}$ & a\\
R41 & $\mathrm{OH}+\mathrm{OH}+\mathrm{M}$ & $\to$ & $\mathrm{H}_{2}\mathrm{O}_{2}+\mathrm{M}$ & $k_{0}=8.97\times10^{-31}(300/T)$, $k_{\infty}=2.6\times 10^{-11}$ & a\\
R42 & $\mathrm{OH}+\mathrm{O}_{3}$ & $\to$ & $\mathrm{HO}_{2}+\mathrm{O}_{2}$ & $1.7\times10^{-12}\exp(-940/T)$ & a\\
R43 & $\mathrm{OH}+\mathrm{HO}_{2}$ & $\to$ & $\mathrm{H}_{2}\mathrm{O}+\mathrm{O}_{2}$ & $4.8\times10^{-11}\exp(250/T)$ & a\\
R44 & $\mathrm{OH}+\mathrm{H}_{2}\mathrm{O}_{2}$ & $\to$ & $\mathrm{H}_{2}\mathrm{O}+\mathrm{HO}_{2}$ & $1.8\times10^{-12}$ & a\\
R45 & $\mathrm{HO}_{2}+\mathrm{O}_{3}$ & $\to$ & $\mathrm{OH}+\mathrm{O}_{2}+\mathrm{O}_{2}$ & $1.0\times10^{-14}\exp(-490/T)$ & a\\
R46 & $\mathrm{HO}_{2}+\mathrm{HO}_{2}$ & $\to$ & $\mathrm{H}_{2}\mathrm{O}_{2}+\mathrm{O}_{2}$ & $3.0\times10^{-13}\exp(460/T)$ & a\\
R47 & $\mathrm{HO}_{2}+\mathrm{HO}_{2}+\mathrm{M}$ & $\to$ & $\mathrm{H}_{2}\mathrm{O}_{2}+\mathrm{O}_{2}+\mathrm{M}$ & $4.2\times10^{-33}\exp(920/T)$ & a\\
R48 & $\mathrm{CO}+\mathrm{OH}+\mathrm{M}$ & $\to$ & $\mathrm{CO}_{2}+\mathrm{H}+\mathrm{M}$ & $k_{0}=1.5\times10^{-13}(300/T)^{-0.6}$, $k_{\infty}=2.1\times 10^{9} (300/T)^{-6.1}$ & a\\
R49 & $\mathrm{CO}+\mathrm{OH}+\mathrm{M}$ & $\to$ & $\mathrm{HOCO}+\mathrm{M}$ & $k_{0}=5.9\times10^{-33}(300/T)^{1.4}$, $k_{\infty}=1.1\times 10^{-12} (300/T)^{-1.3}$ & a\\
R50 & $\mathrm{HOCO}+\mathrm{O}_{2}$ & $\to$ & $\mathrm{HO}_{2}+\mathrm{CO}_{2}$ & $2.0\times10^{-12}$ & a\\
R51 & $\mathrm{CO}_{2}^{+}+\mathrm{H}_{2}$ & $\to$ & $\mathrm{CO}_{2}+\mathrm{H}+\mathrm{H}$ & $8.7\times10^{-10}$ & a\\
R52 & $\mathrm{CO}^{18}\mathrm{O}+h\nu$ & $\to$ & $\mathrm{C}^{18}\mathrm{O}+\mathrm{O}$ & Photodissociation & a\\
R53 & $\mathrm{CO}^{18}\mathrm{O}+h\nu$ & $\to$ & $\mathrm{CO}+{}^{18}\mathrm{O}$ & Photodissociation & a\\
R54 & $\mathrm{CO}^{18}\mathrm{O}+h\nu$ & $\to$ & $\mathrm{C}^{18}\mathrm{O}+\mathrm{O}(^1\mathrm{D})$ & Photodissociation & a\\
R55 & $\mathrm{CO}^{18}\mathrm{O}+h\nu$ & $\to$ & $\mathrm{CO}+{}^{18}\mathrm{O}(^1\mathrm{D})$ & Photodissociation & a\\
R56 & $\mathrm{H}_{2}{}^{18}\mathrm{O}+h\nu$ & $\to$ & $\mathrm{H}+{}^{18}\mathrm{OH}$ & Photodissociation & a\\
R57 & $\mathrm{H}_{2}{}^{18}\mathrm{O}+h\nu$ & $\to$ & $\mathrm{H}_{2}+{}^{18}\mathrm{O}(^1\mathrm{D})$ & Photodissociation & a\\
R58 & $\mathrm{O}^{18}\mathrm{OO}+h\nu$ & $\to$ & $\mathrm{O}^{18}\mathrm{O}+\mathrm{O}$ & Photodissociation & a\\
R59 & $\mathrm{OO}^{18}\mathrm{O}+h\nu$ & $\to$ & $\mathrm{O}_{2}+{}^{18}\mathrm{O}$ & Photodissociation & a\\
R60 & $\mathrm{OO}^{18}\mathrm{O}+h\nu$ & $\to$ & $\mathrm{O}^{18}\mathrm{O}+\mathrm{O}$ & Photodissociation & a\\
R61 & $\mathrm{O}^{18}\mathrm{OO}+h\nu$ & $\to$ & $\mathrm{O}^{18}\mathrm{O}+\mathrm{O}(^1\mathrm{D})$ & Photodissociation & a\\
R62 & $\mathrm{OO}^{18}\mathrm{O}+h\nu$ & $\to$ & $\mathrm{O}_{2}+{}^{18}\mathrm{O}(^1\mathrm{D})$ & Photodissociation & a\\
R63 & $\mathrm{OO}^{18}\mathrm{O}+h\nu$ & $\to$ & $\mathrm{O}^{18}\mathrm{O}+\mathrm{O}(^1\mathrm{D})$ & Photodissociation & a\\
R64 & $\mathrm{O}^{18}\mathrm{O}+h\nu$ & $\to$ & $\mathrm{O}+{}^{18}\mathrm{O}$ & Photodissociation & a\\
R65 & $\mathrm{O}^{18}\mathrm{O}+h\nu$ & $\to$ & $\mathrm{O}+{}^{18}\mathrm{O}(^1\mathrm{D})$ & Photodissociation & a\\
R66 & $\mathrm{O}^{18}\mathrm{O}+h\nu$ & $\to$ & ${}^{18}\mathrm{O}+\mathrm{O}(^1\mathrm{D})$ & Photodissociation & a\\
R67 & ${}^{18}\mathrm{OH}+h\nu$ & $\to$ & ${}^{18}\mathrm{O}+\mathrm{H}$ & Photodissociation & a\\
R68 & ${}^{18}\mathrm{OH}+h\nu$ & $\to$ & ${}^{18}\mathrm{O}(^1\mathrm{D})+\mathrm{H}$ & Photodissociation & a\\
R69 & $\mathrm{HO}^{18}\mathrm{O}+h\nu$ & $\to$ & $\mathrm{OH}+{}^{18}\mathrm{O}$ & Photodissociation & a\\
R70 & $\mathrm{HO}^{18}\mathrm{O}+h\nu$ & $\to$ & ${}^{18}\mathrm{OH}+\mathrm{O}$ & Photodissociation & a\\
R71 & $\mathrm{H}_{2}\mathrm{O}^{18}\mathrm{O}+h\nu$ & $\to$ & $\mathrm{OH}+{}^{18}\mathrm{OH}$ & Photodissociation & a\\
R72 & $\mathrm{H}_{2}\mathrm{O}^{18}\mathrm{O}+h\nu$ & $\to$ & $\mathrm{HO}^{18}\mathrm{O}+\mathrm{H}$ & Photodissociation & a\\
R73 & $\mathrm{H}_{2}\mathrm{O}^{18}\mathrm{O}+h\nu$ & $\to$ & $\mathrm{H}_{2}\mathrm{O}+{}^{18}\mathrm{O}(^1\mathrm{D})$ & Photodissociation & a\\
R74 & $\mathrm{H}_{2}\mathrm{O}^{18}\mathrm{O}+h\nu$ & $\to$ & $\mathrm{H}_{2}{}^{18}\mathrm{O}+\mathrm{O}(^1\mathrm{D})$ & Photodissociation & a\\
R75 & $\mathrm{O}+{}^{18}\mathrm{O}+\mathrm{M}$ & $\to$ & $\mathrm{O}^{18}\mathrm{O}+\mathrm{M}$ & $5.4\times10^{-33}(300/T)^{3.25}\times0.9718$ & a, b\\
R76 & $\mathrm{O}+\mathrm{O}^{18}\mathrm{O}+\mathrm{CO}_{2}$ & $\to$ & $\mathrm{OO}^{18}\mathrm{O}+\mathrm{CO}_{2}$ & $1.5\times10^{-33}(300/T)^{2.4}\times0.5\times(2.0\times10^{-5}(T-295)+1.426)$ & a, c\\
R77 & $\mathrm{O}+\mathrm{O}^{18}\mathrm{O}+\mathrm{CO}_{2}$ & $\to$ & $\mathrm{O}^{18}\mathrm{OO}+\mathrm{CO}_{2}$ & $1.5\times10^{-33}(300/T)^{2.4}\times0.5\times(2.0\times10^{-5}(T-295)+1.007)$ & a, c\\
R78 & ${}^{18}\mathrm{O}+\mathrm{O}_{2}+\mathrm{CO}_{2}$ & $\to$ & $\mathrm{OO}^{18}\mathrm{O}+\mathrm{CO}_{2}$ & $1.5\times10^{-33}(300/T)^{2.4}\times(1.03\times10^{-3}(T-296)+0.92)$ & a, c\\
R79 & $\mathrm{O}+\mathrm{O}^{18}\mathrm{O}$ & $\to$ & ${}^{18}\mathrm{O}+\mathrm{O}_{2}$ & $1.75\times10^{-12}(300/T)^{1.1}\exp(-32.0/T)$ & d\\
R80 & ${}^{18}\mathrm{O}+\mathrm{O}_{2}$ & $\to$ & $\mathrm{O}+\mathrm{O}^{18}\mathrm{O}$ & $3.40\times10^{-12}(300/T)^{1.1}$ & d\\
R81 & $\mathrm{O}+\mathrm{OO}^{18}\mathrm{O}$ & $\to$ & $\mathrm{O}^{18}\mathrm{O}+\mathrm{O}_{2}$ & $8.0\times10^{-12}\exp(-2060/T)\times0.9949$ & a, b\\
R82 & $\mathrm{O}+\mathrm{O}^{18}\mathrm{OO}$ & $\to$ & $\mathrm{O}^{18}\mathrm{O}+\mathrm{O}_{2}$ & $8.0\times10^{-12}\exp(-2060/T)\times0.9949$ & a, b\\
R83 & ${}^{18}\mathrm{O}+\mathrm{O}_{3}$ & $\to$ & $\mathrm{O}^{18}\mathrm{O}+\mathrm{O}_{2}$ & $8.0\times10^{-12}\exp(-2060/T)\times0.9574$ & a, b\\
R84 & $\mathrm{O}+\mathrm{C}^{18}\mathrm{O}+\mathrm{M}$ & $\to$ & $\mathrm{CO}^{18}\mathrm{O}+\mathrm{M}$ & $2.2\times10^{-33}\exp(-1780/T)\times0.9878$ & a, b\\
R85 & ${}^{18}\mathrm{O}+\mathrm{CO}+\mathrm{M}$ & $\to$ & $\mathrm{CO}^{18}\mathrm{O}+\mathrm{M}$ & $2.2\times10^{-33}\exp(-1780/T)\times0.9639$ & a, b\\
R86 & ${}^{18}\mathrm{O}(^1\mathrm{D})+\mathrm{O}_{2}$ & $\to$ & ${}^{18}\mathrm{O}+\mathrm{O}_{2}$ & $3.2\times10^{-11}\exp(70/T)\times0.9622$ & a, b\\
R87 & $\mathrm{O}(^1\mathrm{D})+\mathrm{OO}^{18}\mathrm{O}$ & $\to$ & $\mathrm{O}^{18}\mathrm{O}+\mathrm{O}_{2}$ & $1.2\times10^{-10}\times0.9949$ & a, b\\
R88 & $\mathrm{O}(^1\mathrm{D})+\mathrm{O}^{18}\mathrm{OO}$ & $\to$ & $\mathrm{O}^{18}\mathrm{O}+\mathrm{O}_{2}$ & $1.2\times10^{-10}\times0.9949$ & a, b\\
R89 & ${}^{18}\mathrm{O}(^1\mathrm{D})+\mathrm{O}_{3}$ & $\to$ & $\mathrm{O}^{18}\mathrm{O}+\mathrm{O}_{2}$ & $1.2\times10^{-10}\times0.9574$ & a, b\\
R90 & $\mathrm{O}(^1\mathrm{D})+\mathrm{OO}^{18}\mathrm{O}$ & $\to$ & $\mathrm{O}+\mathrm{O}+\mathrm{O}^{18}\mathrm{O}$ & $0.5\times1.2\times10^{-10}\times0.9949$ & a, b\\
R91 & $\mathrm{O}(^1\mathrm{D})+\mathrm{OO}^{18}\mathrm{O}$ & $\to$ & $\mathrm{O}+{}^{18}\mathrm{O}+\mathrm{O}_{2}$ & $0.5\times1.2\times10^{-10}\times0.9949$ & a, b\\
R92 & $\mathrm{O}(^1\mathrm{D})+\mathrm{O}^{18}\mathrm{OO}$ & $\to$ & $\mathrm{O}+\mathrm{O}+\mathrm{O}^{18}\mathrm{O}$ & $1.2\times10^{-10}\times0.9949$ & a, b\\
R93 & ${}^{18}\mathrm{O}(^1\mathrm{D})+\mathrm{O}_{3}$ & $\to$ & ${}^{18}\mathrm{O}+\mathrm{O}+\mathrm{O}_{2}$ & $1.2\times10^{-10}\times0.9574$ & a, b\\
R94 & ${}^{18}\mathrm{O}(^1\mathrm{D})+\mathrm{H}_{2}$ & $\to$ & $\mathrm{H}+{}^{18}\mathrm{OH}$ & $1.2\times10^{-10}\times0.9938$ & a, b\\
R95 & ${}^{18}\mathrm{O}(^1\mathrm{D})+\mathrm{CO}_{2}$ & $\to$ & ${}^{18}\mathrm{O}+\mathrm{CO}_{2}$ & $(1/3)\times7.5\times10^{-11}\exp(115/T)\times0.9583$ & a, b, d\\
R96 & ${}^{18}\mathrm{O}(^1\mathrm{D})+\mathrm{CO}_{2}$ & $\to$ & $\mathrm{O}+\mathrm{CO}^{18}\mathrm{O}$ & $(2/3)\times7.5\times10^{-11}\exp(115/T)\times0.9583$ & a, b, d\\
R97 & $\mathrm{O}(^1\mathrm{D})+\mathrm{CO}^{18}\mathrm{O}$ & $\to$ & ${}^{18}\mathrm{O}+\mathrm{CO}_{2}$ & $(1/3)\times7.5\times10^{-11}\exp(115/T)\times0.9941$ & a, b, d\\
R98 & $\mathrm{O}(^1\mathrm{D})+\mathrm{CO}^{18}\mathrm{O}$ & $\to$ & $\mathrm{O}+\mathrm{CO}^{18}\mathrm{O}$ & $(2/3)\times7.5\times10^{-11}\exp(115/T)\times0.9941$ & a, b, d\\
R99 & $\mathrm{O}(^1\mathrm{D})+\mathrm{H}_{2}{}^{18}\mathrm{O}$ & $\to$ & $\mathrm{OH}+{}^{18}\mathrm{OH}$ & $1.63\times10^{-10}\exp(60/T)\times0.9761$ & a, b\\
R100 & ${}^{18}\mathrm{O}(^1\mathrm{D})+\mathrm{H}_{2}\mathrm{O}$ & $\to$ & $\mathrm{OH}+{}^{18}\mathrm{OH}$ & $1.63\times10^{-10}\exp(60/T)\times0.9701$ & a, b\\
R101 & $\mathrm{H}_{2}+{}^{18}\mathrm{O}$ & $\to$ & ${}^{18}\mathrm{OH}+\mathrm{H}$ & $6.34\times10^{-12}\exp(-4000/T)\times0.9938$ & a, b\\
R102 & ${}^{18}\mathrm{OH}+\mathrm{H}_{2}$ & $\to$ & $\mathrm{H}_{2}{}^{18}\mathrm{O}+\mathrm{H}$ & $9.01\times10^{-13}\exp(-1526/T)\times0.9944$ & a, b\\
R103 & $\mathrm{H}+{}^{18}\mathrm{OH}+\mathrm{CO}_{2}$ & $\to$ & $\mathrm{H}_{2}{}^{18}\mathrm{O}+\mathrm{CO}_{2}$ & $1.292\times10^{-30}(300/T)^{2}\times0.9970$ & a, b\\
R104 & $\mathrm{H}+\mathrm{HO}^{18}\mathrm{O}$ & $\to$ & $\mathrm{OH}+{}^{18}\mathrm{OH}$ & $7.2\times10^{-11}\times0.9991$ & a, b\\
R105 & $\mathrm{H}+\mathrm{HO}^{18}\mathrm{O}$ & $\to$ & $\mathrm{H}_{2}\mathrm{O}+{}^{18}\mathrm{O}(^1\mathrm{D})$ & $0.5\times1.6\times10^{-12}\times0.9991$ & a, b\\
R106 & $\mathrm{H}+\mathrm{HO}^{18}\mathrm{O}$ & $\to$ & $\mathrm{H}_{2}{}^{18}\mathrm{O}+\mathrm{O}(^1\mathrm{D})$ & $0.5\times1.6\times10^{-12}\times0.9991$ & a, b\\
R107 & $\mathrm{H}+\mathrm{HO}^{18}\mathrm{O}$ & $\to$ & $\mathrm{H}_{2}+\mathrm{O}^{18}\mathrm{O}$ & $3.45\times10^{-12}\times0.9991$ & a, b\\
R108 & $\mathrm{H}+\mathrm{H}_{2}\mathrm{O}^{18}\mathrm{O}$ & $\to$ & $\mathrm{HO}^{18}\mathrm{O}+\mathrm{H}_{2}$ & $2.8\times10^{-12}\exp(-1890/T)\times0.9992$ & a, b\\
R109 & $\mathrm{H}+\mathrm{H}_{2}\mathrm{O}^{18}\mathrm{O}$ & $\to$ & $\mathrm{H}_{2}{}^{18}\mathrm{O}+\mathrm{OH}$ & $0.5\times1.7\times10^{-11}\exp(-1800/T)\times0.9992$ & a, b\\
R110 & $\mathrm{H}+\mathrm{H}_{2}\mathrm{O}^{18}\mathrm{O}$ & $\to$ & $\mathrm{H}_{2}\mathrm{O}+{}^{18}\mathrm{OH}$ & $0.5\times1.7\times10^{-11}\exp(-1800/T)\times0.9992$ & a, b\\
R111 & $\mathrm{H}+\mathrm{O}^{18}\mathrm{O}+\mathrm{M}$ & $\to$ & $\mathrm{HO}^{18}\mathrm{O}+\mathrm{M}$ & $k_{0}=8.8\times10^{-32}(300/T)^{1.3}\times0.9991$, $k_{\infty}=7.5\times 10^{-11} (300/T)^{-0.2}\times 0.9991$ & a, b\\
R112 & $\mathrm{H}+\mathrm{OO}^{18}\mathrm{O}$ & $\to$ & $\mathrm{OH}+\mathrm{O}^{18}\mathrm{O}$ & $0.5\times1.4\times10^{-10}\exp(-470/T)\times0.9995$ & a, b\\
R113 & $\mathrm{H}+\mathrm{OO}^{18}\mathrm{O}$ & $\to$ & ${}^{18}\mathrm{OH}+\mathrm{O}_{2}$ & $0.5\times1.4\times10^{-10}\exp(-470/T)\times0.9995$ & a, b\\
R114 & $\mathrm{H}+\mathrm{O}^{18}\mathrm{OO}$ & $\to$ & $\mathrm{OH}+\mathrm{O}^{18}\mathrm{O}$ & $1.4\times10^{-10}\exp(-470/T)\times0.9995$ & a, b\\
R115 & $\mathrm{O}+{}^{18}\mathrm{OH}$ & $\to$ & $\mathrm{O}^{18}\mathrm{O}+\mathrm{H}$ & $1.8\times10^{-11}\exp(180/T)\times0.9741$ & a, b\\
R116 & ${}^{18}\mathrm{O}+\mathrm{OH}$ & $\to$ & $\mathrm{O}^{18}\mathrm{O}+\mathrm{H}$ & $1.8\times10^{-11}\exp(180/T)\times0.9709$ & a, b\\
R117 & $\mathrm{O}+\mathrm{HO}^{18}\mathrm{O}$ & $\to$ & $\mathrm{OH}+\mathrm{O}^{18}\mathrm{O}$ & $0.5\times3.0\times10^{-11}\exp(200/T)\times0.9906$ & a, b\\
R118 & $\mathrm{O}+\mathrm{HO}^{18}\mathrm{O}$ & $\to$ & ${}^{18}\mathrm{OH}+\mathrm{O}_{2}$ & $0.5\times3.0\times10^{-11}\exp(200/T)\times0.9906$ & a, b\\
R119 & ${}^{18}\mathrm{O}+\mathrm{HO}_{2}$ & $\to$ & $\mathrm{OH}+\mathrm{O}^{18}\mathrm{O}$ & $3.0\times10^{-11}\exp(200/T)\times0.9618$ & a, b\\
R120 & $\mathrm{O}+\mathrm{H}_{2}\mathrm{O}^{18}\mathrm{O}$ & $\to$ & $\mathrm{OH}+\mathrm{HO}^{18}\mathrm{O}$ & $0.5\times1.4\times10^{-12}\exp(-2000/T)\times0.9910$ & a, b\\
R121 & $\mathrm{O}+\mathrm{H}_{2}\mathrm{O}^{18}\mathrm{O}$ & $\to$ & ${}^{18}\mathrm{OH}+\mathrm{HO}_{2}$ & $0.5\times1.4\times10^{-12}\exp(-2000/T)\times0.9910$ & a, b\\
R122 & ${}^{18}\mathrm{O}+\mathrm{H}_{2}\mathrm{O}_{2}$ & $\to$ & $\mathrm{OH}+\mathrm{HO}^{18}\mathrm{O}$ & $0.5\times1.4\times10^{-12}\exp(-2000/T)\times0.9614$ & a, b\\
R123 & ${}^{18}\mathrm{O}+\mathrm{H}_{2}\mathrm{O}_{2}$ & $\to$ & ${}^{18}\mathrm{OH}+\mathrm{HO}_{2}$ & $0.5\times1.4\times10^{-12}\exp(-2000/T)\times0.9614$ & a, b\\
R124 & ${}^{18}\mathrm{OH}+\mathrm{OH}$ & $\to$ & $\mathrm{H}_{2}{}^{18}\mathrm{O}+\mathrm{O}$ & $2.0\times0.5\times1.8\times10^{-12}\times0.9733$ & a, b\\
R125 & ${}^{18}\mathrm{OH}+\mathrm{OH}$ & $\to$ & $\mathrm{H}_{2}\mathrm{O}+{}^{18}\mathrm{O}$ & $2.0\times0.5\times1.8\times10^{-12}\times0.9733$ & a, b\\
R126 & ${}^{18}\mathrm{OH}+\mathrm{OH}+\mathrm{M}$ & $\to$ & $\mathrm{H}_{2}\mathrm{O}^{18}\mathrm{O}+\mathrm{M}$ & $k_{0}=2.0\times8.97\times10^{-31}(300/T)\times0.9733$, $k_{\infty}=2.6\times 10^{-11}\times0.9733$ & a, b\\
R127 & ${}^{18}\mathrm{OH}+\mathrm{O}_{3}$ & $\to$ & $\mathrm{HO}^{18}\mathrm{O}+\mathrm{O}_{2}$ & $1.7\times10^{-12}\exp(-940/T)\times0.9603$ & a, b\\
R128 & $\mathrm{OH}+\mathrm{OO}^{18}\mathrm{O}$ & $\to$ & $\mathrm{HO}_{2}+\mathrm{O}^{18}\mathrm{O}$ & $0.5\times1.7\times10^{-12}\exp(-940/T)\times0.9947$ & a, b\\
R129 & $\mathrm{OH}+\mathrm{OO}^{18}\mathrm{O}$ & $\to$ & $\mathrm{HO}^{18}\mathrm{O}+\mathrm{O}_{2}$ & $0.5\times1.7\times10^{-12}\exp(-940/T)\times0.9947$ & a, b\\
R130 & $\mathrm{OH}+\mathrm{O}^{18}\mathrm{OO}$ & $\to$ & $\mathrm{HO}_{2}+\mathrm{O}^{18}\mathrm{O}$ & $1.7\times10^{-12}\exp(-940/T)\times0.9947$ & a, b\\
R131 & $\mathrm{HO}_{2}+\mathrm{OO}^{18}\mathrm{O}$ & $\to$ & $\mathrm{OH}+\mathrm{O}_{2}+\mathrm{O}^{18}\mathrm{O}$ & $1.0\times10^{-14}\exp(-490/T)\times0.9918$ & a, b\\
R132 & $\mathrm{HO}_{2}+\mathrm{O}^{18}\mathrm{OO}$ & $\to$ & $\mathrm{OH}+\mathrm{O}_{2}+\mathrm{O}^{18}\mathrm{O}$ & $1.0\times10^{-14}\exp(-490/T)\times0.9918$ & a, b\\
R133 & ${}^{18}\mathrm{OH}+\mathrm{HO}_{2}$ & $\to$ & $\mathrm{H}_{2}{}^{18}\mathrm{O}+\mathrm{O}_{2}$ & $4.8\times10^{-11}\exp(250/T)\times0.9646$ & a, b\\
R134 & $\mathrm{OH}+\mathrm{HO}^{18}\mathrm{O}$ & $\to$ & $\mathrm{H}_{2}\mathrm{O}+\mathrm{O}^{18}\mathrm{O}$ & $4.8\times10^{-11}\exp(250/T)\times0.9902$ & a, b\\
R135 & $\mathrm{OH}+\mathrm{H}_{2}\mathrm{O}^{18}\mathrm{O}$ & $\to$ & $\mathrm{H}_{2}\mathrm{O}+\mathrm{HO}^{18}\mathrm{O}$ & $1.8\times10^{-12}\times0.9906$ & a, b\\
R136 & ${}^{18}\mathrm{OH}+\mathrm{H}_{2}\mathrm{O}_{2}$ & $\to$ & $\mathrm{H}_{2}{}^{18}\mathrm{O}+\mathrm{HO}_{2}$ & $1.8\times10^{-12}\times0.9642$ & a, b\\
R137 & $\mathrm{HO}_{2}+\mathrm{HO}^{18}\mathrm{O}$ & $\to$ & $\mathrm{H}_{2}\mathrm{O}^{18}\mathrm{O}+\mathrm{O}_{2}$ & $2.0\times0.5\times3.0\times10^{-13}\exp(460/T)\times0.9856$ & a, b\\
R138 & $\mathrm{HO}_{2}+\mathrm{HO}^{18}\mathrm{O}$ & $\to$ & $\mathrm{H}_{2}\mathrm{O}_{2}+\mathrm{O}^{18}\mathrm{O}$ & $2.0\times0.5\times3.0\times10^{-13}\exp(460/T)\times0.9856$ & a, b\\
R139 & $\mathrm{HO}_{2}+\mathrm{HO}^{18}\mathrm{O}+\mathrm{M}$ & $\to$ & $\mathrm{H}_{2}\mathrm{O}^{18}\mathrm{O}+\mathrm{O}_{2}+\mathrm{M}$ & $2.0\times0.5\times4.2\times10^{-33}\exp(920/T)\times0.9856$ & a, b\\
R140 & $\mathrm{HO}_{2}+\mathrm{HO}^{18}\mathrm{O}+\mathrm{M}$ & $\to$ & $\mathrm{H}_{2}\mathrm{O}_{2}+\mathrm{O}^{18}\mathrm{O}+\mathrm{M}$ & $2.0\times0.5\times4.2\times10^{-33}\exp(920/T)\times0.9856$ & a, b\\
R141 & $\mathrm{C}^{18}\mathrm{O}+\mathrm{OH}+\mathrm{M}$ & $\to$ & $\mathrm{CO}^{18}\mathrm{O}+\mathrm{H}+\mathrm{M}$ & $k_{0}=1.5\times10^{-13}(300/T)^{-0.6}\times0.9873$, $k_{\infty}=2.1\times 10^{9} (300/T)^{-6.1}\times0.9873$  & a, b\\
R142 & $\mathrm{CO}+{}^{18}\mathrm{OH}+\mathrm{M}$ & $\to$ & $\mathrm{CO}^{18}\mathrm{O}+\mathrm{H}+\mathrm{M}$ & $k_{0}=1.5\times10^{-13}(300/T)^{-0.6}\times0.9666$, $k_{\infty}=2.1\times 10^{9} (300/T)^{-6.1}\times0.9666$  & a, b\\
R143 & $\mathrm{C}^{18}\mathrm{O}+\mathrm{OH}+\mathrm{M}$ & $\to$ & $\mathrm{HOC}^{18}\mathrm{O}+\mathrm{M}$ & $k_{0}=5.9\times10^{-33}(300/T)^{1.4}\times0.9873$, $k_{\infty}=1.1\times 10^{-12} (300/T)^{-1.3}\times0.9873$ & a, b\\
R144 & $\mathrm{CO}+{}^{18}\mathrm{OH}+\mathrm{M}$ & $\to$ & $\mathrm{H}^{18}\mathrm{OCO}+\mathrm{M}$ & $k_{0}=5.9\times10^{-33}(300/T)^{1.4}\times0.9666$, $k_{\infty}=1.1\times 10^{-12} (300/T)^{-1.3}\times0.9666$ & a, b\\
R145 & $\mathrm{HOC}^{18}\mathrm{O}+\mathrm{O}_{2}$ & $\to$ & $\mathrm{HO}_{2}+\mathrm{CO}^{18}\mathrm{O}$ & $2.0\times10^{-12}\times0.9911$ & a, b\\
R146 & $\mathrm{H}^{18}\mathrm{OCO}+\mathrm{O}_{2}$ & $\to$ & $\mathrm{HO}^{18}\mathrm{O}+\mathrm{CO}_{2}$ & $2.0\times10^{-12}\times0.9911$ & a, b\\
R147 & $\mathrm{HOCO}+\mathrm{O}^{18}\mathrm{O}$ & $\to$ & $\mathrm{HO}_{2}+\mathrm{CO}^{18}\mathrm{O}$ & $0.5\times2.0\times10^{-12}\times0.9826$ & a, b\\
R148 & $\mathrm{HOCO}+\mathrm{O}^{18}\mathrm{O}$ & $\to$ & $\mathrm{HO}^{18}\mathrm{O}+\mathrm{CO}_{2}$ & $0.5\times2.0\times10^{-12}\times0.9826$ & a, b\\
R149 & $\mathrm{CO}^{17}\mathrm{O}+h\nu$ & $\to$ & $\mathrm{C}^{17}\mathrm{O}+\mathrm{O}$ & Photodissociation & a\\
R150 & $\mathrm{CO}^{17}\mathrm{O}+h\nu$ & $\to$ & $\mathrm{CO}+{}^{17}\mathrm{O}$ & Photodissociation & a\\
R151 & $\mathrm{CO}^{17}\mathrm{O}+h\nu$ & $\to$ & $\mathrm{C}^{17}\mathrm{O}+\mathrm{O}(^1\mathrm{D})$ & Photodissociation & a\\
R152 & $\mathrm{CO}^{17}\mathrm{O}+h\nu$ & $\to$ & $\mathrm{CO}+{}^{17}\mathrm{O}(^1\mathrm{D})$ & Photodissociation & a\\
R153 & $\mathrm{H}_{2}{}^{17}\mathrm{O}+h\nu$ & $\to$ & $\mathrm{H}+{}^{17}\mathrm{OH}$ & Photodissociation & a\\
R154 & $\mathrm{H}_{2}{}^{17}\mathrm{O}+h\nu$ & $\to$ & $\mathrm{H}_{2}+{}^{17}\mathrm{O}(^1\mathrm{D})$ & Photodissociation & a\\
R155 & $\mathrm{O}^{17}\mathrm{OO}+h\nu$ & $\to$ & $\mathrm{O}^{17}\mathrm{O}+\mathrm{O}$ & Photodissociation & a\\
R156 & $\mathrm{OO}^{17}\mathrm{O}+h\nu$ & $\to$ & $\mathrm{O}_{2}+{}^{17}\mathrm{O}$ & Photodissociation & a\\
R157 & $\mathrm{OO}^{17}\mathrm{O}+h\nu$ & $\to$ & $\mathrm{O}^{17}\mathrm{O}+\mathrm{O}$ & Photodissociation & a\\
R158 & $\mathrm{O}^{17}\mathrm{OO}+h\nu$ & $\to$ & $\mathrm{O}^{17}\mathrm{O}+\mathrm{O}(^1\mathrm{D})$ & Photodissociation & a\\
R159 & $\mathrm{OO}^{17}\mathrm{O}+h\nu$ & $\to$ & $\mathrm{O}_{2}+{}^{17}\mathrm{O}(^1\mathrm{D})$ & Photodissociation & a\\
R160 & $\mathrm{OO}^{17}\mathrm{O}+h\nu$ & $\to$ & $\mathrm{O}^{17}\mathrm{O}+\mathrm{O}(^1\mathrm{D})$ & Photodissociation & a\\
R161 & $\mathrm{O}^{17}\mathrm{O}+h\nu$ & $\to$ & $\mathrm{O}+{}^{17}\mathrm{O}$ & Photodissociation & a\\
R162 & $\mathrm{O}^{17}\mathrm{O}+h\nu$ & $\to$ & $\mathrm{O}+{}^{17}\mathrm{O}(^1\mathrm{D})$ & Photodissociation & a\\
R163 & $\mathrm{O}^{17}\mathrm{O}+h\nu$ & $\to$ & ${}^{17}\mathrm{O}+\mathrm{O}(^1\mathrm{D})$ & Photodissociation & a\\
R164 & ${}^{17}\mathrm{OH}+h\nu$ & $\to$ & ${}^{17}\mathrm{O}+\mathrm{H}$ & Photodissociation & a\\
R165 & ${}^{17}\mathrm{OH}+h\nu$ & $\to$ & ${}^{17}\mathrm{O}(^1\mathrm{D})+\mathrm{H}$ & Photodissociation & a\\
R166 & $\mathrm{HO}^{17}\mathrm{O}+h\nu$ & $\to$ & $\mathrm{OH}+{}^{17}\mathrm{O}$ & Photodissociation & a\\
R167 & $\mathrm{HO}^{17}\mathrm{O}+h\nu$ & $\to$ & ${}^{17}\mathrm{OH}+\mathrm{O}$ & Photodissociation & a\\
R168 & $\mathrm{H}_{2}\mathrm{O}^{17}\mathrm{O}+h\nu$ & $\to$ & $\mathrm{OH}+{}^{17}\mathrm{OH}$ & Photodissociation & a\\
R169 & $\mathrm{H}_{2}\mathrm{O}^{17}\mathrm{O}+h\nu$ & $\to$ & $\mathrm{HO}^{17}\mathrm{O}+\mathrm{H}$ & Photodissociation & a\\
R170 & $\mathrm{H}_{2}\mathrm{O}^{17}\mathrm{O}+h\nu$ & $\to$ & $\mathrm{H}_{2}\mathrm{O}+{}^{17}\mathrm{O}(^1\mathrm{D})$ & Photodissociation & a\\
R171 & $\mathrm{H}_{2}\mathrm{O}^{17}\mathrm{O}+h\nu$ & $\to$ & $\mathrm{H}_{2}{}^{17}\mathrm{O}+\mathrm{O}(^1\mathrm{D})$ & Photodissociation & a\\
R172 & $\mathrm{O}+{}^{17}\mathrm{O}+\mathrm{M}$ & $\to$ & $\mathrm{O}^{17}\mathrm{O}+\mathrm{M}$ & $5.4\times10^{-33}(300/T)^{3.25} \times 0.9851$ & a, b\\
R173 & $\mathrm{O}+\mathrm{O}^{17}\mathrm{O}+\mathrm{CO}_{2}$ & $\to$ & $\mathrm{OO}^{17}\mathrm{O}+\mathrm{CO}_{2}$ & $1.5\times10^{-33}(300/T)^{2.4}\times0.5\times(1.477\times10^{-5}(T-295)+1.292)$ & a, c\\
R174 & $\mathrm{O}+\mathrm{O}^{17}\mathrm{O}+\mathrm{CO}_{2}$ & $\to$ & $\mathrm{O}^{17}\mathrm{OO}+\mathrm{CO}_{2}$ & $1.5\times10^{-33}(300/T)^{2.4}\times0.5\times(1.477\times10^{-5}(T-295)+1.005)$ & a, c\\
R175 & ${}^{17}\mathrm{O}+\mathrm{O}_{2}+\mathrm{CO}_{2}$ & $\to$ & $\mathrm{OO}^{17}\mathrm{O}+\mathrm{CO}_{2}$ & $1.5\times10^{-33}(300/T)^{2.4}\times(7.606\times10^{-4}(T-296)+1.03)$ & a, c\\
R176 & $\mathrm{O}+\mathrm{O}^{17}\mathrm{O}$ & $\to$ & ${}^{17}\mathrm{O}+\mathrm{O}_{2}$ & $1.73\times10^{-12}(300/T)^{1.1}\exp(-16.8/T)$ & d\\
R177 & ${}^{17}\mathrm{O}+\mathrm{O}_{2}$ & $\to$ & $\mathrm{O}+\mathrm{O}^{17}\mathrm{O}$ & $3.4\times10^{-12}(300/T)^{1.1}$ & d\\
R178 & $\mathrm{O}+\mathrm{OO}^{17}\mathrm{O}$ & $\to$ & $\mathrm{O}^{17}\mathrm{O}+\mathrm{O}_{2}$ & $8.0\times10^{-12}\exp(-2060/T)\times0.9974$ & a, b\\
R179 & $\mathrm{O}+\mathrm{O}^{17}\mathrm{OO}$ & $\to$ & $\mathrm{O}^{17}\mathrm{O}+\mathrm{O}_{2}$ & $8.0\times10^{-12}\exp(-2060/T)\times0.9974$ & a, b\\
R180 & ${}^{17}\mathrm{O}+\mathrm{O}_{3}$ & $\to$ & $\mathrm{O}^{17}\mathrm{O}+\mathrm{O}_{2}$ & $8.0\times10^{-12}\exp(-2060/T)\times0.9776$ & a, b\\
R181 & $\mathrm{O}+\mathrm{C}^{17}\mathrm{O}+\mathrm{M}$ & $\to$ & $\mathrm{CO}^{17}\mathrm{O}+\mathrm{M}$ & $2.2\times10^{-33}\exp(-1780/T)\times0.9937$ & a, b\\
R182 & ${}^{17}\mathrm{O}+\mathrm{CO}+\mathrm{M}$ & $\to$ & $\mathrm{CO}^{17}\mathrm{O}+\mathrm{M}$ & $2.2\times10^{-33}\exp(-1780/T)\times0.9811$ & a, b\\
R183 & ${}^{17}\mathrm{O}(^1\mathrm{D})+\mathrm{O}_{2}$ & $\to$ & ${}^{17}\mathrm{O}+\mathrm{O}_{2}$ & $3.2\times10^{-11}\exp(70/T)\times0.9801$ & a, b\\
R184 & $\mathrm{O}(^1\mathrm{D})+\mathrm{OO}^{17}\mathrm{O}$ & $\to$ & $\mathrm{O}^{17}\mathrm{O}+\mathrm{O}_{2}$ & $1.2\times10^{-10}\times0.9974$ & a, b\\
R185 & $\mathrm{O}(^1\mathrm{D})+\mathrm{O}^{17}\mathrm{OO}$ & $\to$ & $\mathrm{O}^{17}\mathrm{O}+\mathrm{O}_{2}$ & $1.2\times10^{-10}\times0.9974$ & a, b\\
R186 & ${}^{17}\mathrm{O}(^1\mathrm{D})+\mathrm{O}_{3}$ & $\to$ & $\mathrm{O}^{17}\mathrm{O}+\mathrm{O}_{2}$ & $1.2\times10^{-10}\times0.9776$ & a, b\\
R187 & $\mathrm{O}(^1\mathrm{D})+\mathrm{OO}^{17}\mathrm{O}$ & $\to$ & $\mathrm{O}+\mathrm{O}+\mathrm{O}^{17}\mathrm{O}$ & $0.5\times1.2\times10^{-10}\times0.9974$ & a, b\\
R188 & $\mathrm{O}(^1\mathrm{D})+\mathrm{OO}^{17}\mathrm{O}$ & $\to$ & $\mathrm{O}+{}^{17}\mathrm{O}+\mathrm{O}_{2}$ & $0.5\times1.2\times10^{-10}\times0.9974$ & a, b\\
R189 & $\mathrm{O}(^1\mathrm{D})+\mathrm{O}^{17}\mathrm{OO}$ & $\to$ & $\mathrm{O}+\mathrm{O}+\mathrm{O}^{17}\mathrm{O}$ & $1.2\times10^{-10}\times0.9974$ & a, b\\
R190 & ${}^{17}\mathrm{O}(^1\mathrm{D})+\mathrm{O}_{3}$ & $\to$ & ${}^{17}\mathrm{O}+\mathrm{O}+\mathrm{O}_{2}$ & $1.2\times10^{-10}\times0.9776$ & a, b\\
R191 & ${}^{17}\mathrm{O}(^1\mathrm{D})+\mathrm{H}_{2}$ & $\to$ & $\mathrm{H}+{}^{17}\mathrm{OH}$ & $1.2\times10^{-10}\times0.9967$ & a, b\\
R192 & ${}^{17}\mathrm{O}(^1\mathrm{D})+\mathrm{CO}_{2}$ & $\to$ & ${}^{17}\mathrm{O}+\mathrm{CO}_{2}$ & $(1/3)\times7.5\times10^{-11}\exp(115/T)\times0.9781$ & a, b, d\\
R193 & ${}^{17}\mathrm{O}(^1\mathrm{D})+\mathrm{CO}_{2}$ & $\to$ & $\mathrm{O}+\mathrm{CO}^{17}\mathrm{O}$ & $(2/3)\times7.5\times10^{-11}\exp(115/T)\times0.9781$ & a, b, d\\
R194 & $\mathrm{O}(^1\mathrm{D})+\mathrm{CO}^{17}\mathrm{O}$ & $\to$ & ${}^{17}\mathrm{O}+\mathrm{CO}_{2}$ & $(1/3)\times7.5\times10^{-11}\exp(115/T)\times0.9970$ & a, b, d\\
R195 & $\mathrm{O}(^1\mathrm{D})+\mathrm{CO}^{17}\mathrm{O}$ & $\to$ & $\mathrm{O}+\mathrm{CO}^{17}\mathrm{O}$ & $(2/3)\times7.5\times10^{-11}\exp(115/T)\times0.9970$ & a, b, d\\
R196 & $\mathrm{O}(^1\mathrm{D})+\mathrm{H}_{2}{}^{17}\mathrm{O}$ & $\to$ & $\mathrm{OH}+{}^{17}\mathrm{OH}$ & $1.63\times10^{-10}\exp(60/T)\times0.9875$ & a, b\\
R197 & ${}^{17}\mathrm{O}(^1\mathrm{D})+\mathrm{H}_{2}\mathrm{O}$ & $\to$ & $\mathrm{OH}+{}^{17}\mathrm{OH}$ & $1.63\times10^{-10}\exp(60/T)\times0.9843$ & a, b\\
R198 & $\mathrm{H}_{2}+{}^{17}\mathrm{O}$ & $\to$ & ${}^{17}\mathrm{OH}+\mathrm{H}$ & $6.34\times10^{-12}\exp(-4000/T)\times0.9967$ & a, b\\
R199 & ${}^{17}\mathrm{OH}+\mathrm{H}_{2}$ & $\to$ & $\mathrm{H}_{2}{}^{17}\mathrm{O}+\mathrm{H}$ & $9.01\times10^{-13}\exp(-1526/T)\times0.9970$ & a, b\\
R200 & $\mathrm{H}+{}^{17}\mathrm{OH}+\mathrm{CO}_{2}$ & $\to$ & $\mathrm{H}_{2}{}^{17}\mathrm{O}+\mathrm{CO}_{2}$ & $1.292\times10^{-30}(300/T)^{2}\times0.9984$ & a, b\\
R201 & $\mathrm{H}+\mathrm{HO}^{17}\mathrm{O}$ & $\to$ & $\mathrm{OH}+{}^{17}\mathrm{OH}$ & $7.2\times10^{-11}\times0.9995$ & a, b\\
R202 & $\mathrm{H}+\mathrm{HO}^{17}\mathrm{O}$ & $\to$ & $\mathrm{H}_{2}\mathrm{O}+{}^{17}\mathrm{O}(^1\mathrm{D})$ & $0.5\times1.6\times10^{-12}\times0.9995$ & a, b\\
R203 & $\mathrm{H}+\mathrm{HO}^{17}\mathrm{O}$ & $\to$ & $\mathrm{H}_{2}{}^{17}\mathrm{O}+\mathrm{O}(^1\mathrm{D})$ & $0.5\times1.6\times10^{-12}\times0.9995$ & a, b\\
R204 & $\mathrm{H}+\mathrm{HO}^{17}\mathrm{O}$ & $\to$ & $\mathrm{H}_{2}+\mathrm{O}^{17}\mathrm{O}$ & $3.45\times10^{-12}\times0.9995$ & a, b\\
R205 & $\mathrm{H}+\mathrm{H}_{2}\mathrm{O}^{17}\mathrm{O}$ & $\to$ & $\mathrm{HO}^{17}\mathrm{O}+\mathrm{H}_{2}$ & $2.8\times10^{-12}\exp(-1890/T)\times0.9996$ & a, b\\
R206 & $\mathrm{H}+\mathrm{H}_{2}\mathrm{O}^{17}\mathrm{O}$ & $\to$ & $\mathrm{H}_{2}{}^{17}\mathrm{O}+\mathrm{OH}$ & $0.5\times1.7\times10^{-11}\exp(-1800/T)\times0.9996$ & a, b\\
R207 & $\mathrm{H}+\mathrm{H}_{2}\mathrm{O}^{17}\mathrm{O}$ & $\to$ & $\mathrm{H}_{2}\mathrm{O}+{}^{17}\mathrm{OH}$ & $0.5\times1.7\times10^{-11}\exp(-1800/T)\times0.9996$ & a, b\\
R208 & $\mathrm{H}+\mathrm{O}^{17}\mathrm{O}+\mathrm{M}$ & $\to$ & $\mathrm{HO}^{17}\mathrm{O}+\mathrm{M}$ & $k_{0}=8.8\times10^{-32}(300/T)^{1.3}\times0.9995$, $k_{\infty}=7.5\times 10^{-11} (300/T)^{-0.2}\times0.9995$ & a, b\\
R209 & $\mathrm{H}+\mathrm{OO}^{17}\mathrm{O}$ & $\to$ & $\mathrm{OH}+\mathrm{O}^{17}\mathrm{O}$ & $0.5\times1.4\times10^{-10}\exp(-470/T)\times0.9998$ & a, b\\
R210 & $\mathrm{H}+\mathrm{OO}^{17}\mathrm{O}$ & $\to$ & ${}^{17}\mathrm{OH}+\mathrm{O}_{2}$ & $0.5\times1.4\times10^{-10}\exp(-470/T)\times0.9998$ & a, b\\
R211 & $\mathrm{H}+\mathrm{O}^{17}\mathrm{OO}$ & $\to$ & $\mathrm{OH}+\mathrm{O}^{17}\mathrm{O}$ & $1.4\times10^{-10}\exp(-470/T)\times0.9998$ & a, b\\
R212 & $\mathrm{O}+{}^{17}\mathrm{OH}$ & $\to$ & $\mathrm{O}^{17}\mathrm{O}+\mathrm{H}$ & $1.8\times10^{-11}\exp(180/T)\times0.9864$ & a, b\\
R213 & ${}^{17}\mathrm{O}+\mathrm{OH}$ & $\to$ & $\mathrm{O}^{17}\mathrm{O}+\mathrm{H}$ & $1.8\times10^{-11}\exp(180/T)\times0.9847$ & a, b\\
R214 & $\mathrm{O}+\mathrm{HO}^{17}\mathrm{O}$ & $\to$ & $\mathrm{OH}+\mathrm{O}^{17}\mathrm{O}$ & $0.5\times3.0\times10^{-11}\exp(200/T)\times0.9951$ & a, b\\
R215 & $\mathrm{O}+\mathrm{HO}^{17}\mathrm{O}$ & $\to$ & ${}^{17}\mathrm{OH}+\mathrm{O}_{2}$ & $0.5\times3.0\times10^{-11}\exp(200/T)\times0.9951$ & a, b\\
R216 & ${}^{17}\mathrm{O}+\mathrm{HO}_{2}$ & $\to$ & $\mathrm{OH}+\mathrm{O}^{17}\mathrm{O}$ & $3.0\times10^{-11}\exp(200/T)\times0.9799$ & a, b\\
R217 & $\mathrm{O}+\mathrm{H}_{2}\mathrm{O}^{17}\mathrm{O}$ & $\to$ & $\mathrm{OH}+\mathrm{HO}^{17}\mathrm{O}$ & $0.5\times1.4\times10^{-12}\exp(-2000/T)\times0.9954$ & a, b\\
R218 & $\mathrm{O}+\mathrm{H}_{2}\mathrm{O}^{17}\mathrm{O}$ & $\to$ & ${}^{17}\mathrm{OH}+\mathrm{HO}_{2}$ & $0.5\times1.4\times10^{-12}\exp(-2000/T)\times0.9954$ & a, b\\
R219 & ${}^{17}\mathrm{O}+\mathrm{H}_{2}\mathrm{O}_{2}$ & $\to$ & $\mathrm{OH}+\mathrm{HO}^{17}\mathrm{O}$ & $0.5\times1.4\times10^{-12}\exp(-2000/T)\times0.9798$ & a, b\\
R220 & ${}^{17}\mathrm{O}+\mathrm{H}_{2}\mathrm{O}_{2}$ & $\to$ & ${}^{17}\mathrm{OH}+\mathrm{HO}_{2}$ & $0.5\times1.4\times10^{-12}\exp(-2000/T)\times0.9798$ & a, b\\
R221 & ${}^{17}\mathrm{OH}+\mathrm{OH}$ & $\to$ & $\mathrm{H}_{2}{}^{17}\mathrm{O}+\mathrm{O}$ & $2.0\times0.5\times1.8\times10^{-12}\times0.9860$ & a, b\\
R222 & ${}^{17}\mathrm{OH}+\mathrm{OH}$ & $\to$ & $\mathrm{H}_{2}\mathrm{O}+{}^{17}\mathrm{O}$ & $2.0\times0.5\times1.8\times10^{-12}\times0.9860$ & a, b\\
R223 & ${}^{17}\mathrm{OH}+\mathrm{OH}+\mathrm{M}$ & $\to$ & $\mathrm{H}_{2}\mathrm{O}^{17}\mathrm{O}+\mathrm{M}$ & $k_{0}=2.0\times8.97\times10^{-31}(300/T)\times0.9860$, $k_{\infty}=2.6\times 10^{-11}\times0.9860$ & a, b\\
R224 & ${}^{17}\mathrm{OH}+\mathrm{O}_{3}$ & $\to$ & $\mathrm{HO}^{17}\mathrm{O}+\mathrm{O}_{2}$ & $1.7\times10^{-12}\exp(-940/T)\times0.9793$ & a, b\\
R225 & $\mathrm{OH}+\mathrm{OO}^{17}\mathrm{O}$ & $\to$ & $\mathrm{HO}_{2}+\mathrm{O}^{17}\mathrm{O}$ & $0.5\times1.7\times10^{-12}\exp(-940/T)\times0.9973$ & a, b\\
R226 & $\mathrm{OH}+\mathrm{OO}^{17}\mathrm{O}$ & $\to$ & $\mathrm{HO}^{17}\mathrm{O}+\mathrm{O}_{2}$ & $0.5\times1.7\times10^{-12}\exp(-940/T)\times0.9973$ & a, b\\
R227 & $\mathrm{OH}+\mathrm{O}^{17}\mathrm{OO}$ & $\to$ & $\mathrm{HO}_{2}+\mathrm{O}^{17}\mathrm{O}$ & $1.7\times10^{-12}\exp(-940/T)\times0.9973$ & a, b\\
R228 & $\mathrm{HO}_{2}+\mathrm{OO}^{17}\mathrm{O}$ & $\to$ & $\mathrm{OH}+\mathrm{O}_{2}+\mathrm{O}^{17}\mathrm{O}$ & $1.0\times10^{-14}\exp(-490/T)\times0.9958$ & a, b\\
R229 & $\mathrm{HO}_{2}+\mathrm{O}^{17}\mathrm{OO}$ & $\to$ & $\mathrm{OH}+\mathrm{O}_{2}+\mathrm{O}^{17}\mathrm{O}$ & $1.0\times10^{-14}\exp(-490/T)\times0.9958$ & a, b\\
R230 & ${}^{17}\mathrm{OH}+\mathrm{HO}_{2}$ & $\to$ & $\mathrm{H}_{2}{}^{17}\mathrm{O}+\mathrm{O}_{2}$ & $4.8\times10^{-11}\exp(250/T)\times0.9815$ & a, b\\
R231 & $\mathrm{OH}+\mathrm{HO}^{17}\mathrm{O}$ & $\to$ & $\mathrm{H}_{2}\mathrm{O}+\mathrm{O}^{17}\mathrm{O}$ & $4.8\times10^{-11}\exp(250/T)\times0.9950$ & a, b\\
R232 & $\mathrm{OH}+\mathrm{H}_{2}\mathrm{O}^{17}\mathrm{O}$ & $\to$ & $\mathrm{H}_{2}\mathrm{O}+\mathrm{HO}^{17}\mathrm{O}$ & $1.8\times10^{-12}\times0.9952$ & a, b\\
R233 & ${}^{17}\mathrm{OH}+\mathrm{H}_{2}\mathrm{O}_{2}$ & $\to$ & $\mathrm{H}_{2}{}^{17}\mathrm{O}+\mathrm{HO}_{2}$ & $1.8\times10^{-12}\times0.9813$ & a, b\\
R234 & $\mathrm{HO}_{2}+\mathrm{HO}^{17}\mathrm{O}$ & $\to$ & $\mathrm{H}_{2}\mathrm{O}^{17}\mathrm{O}+\mathrm{O}_{2}$ & $2.0\times0.5\times3.0\times10^{-13}\exp(460/T)\times0.9926$ & a, b\\
R235 & $\mathrm{HO}_{2}+\mathrm{HO}^{17}\mathrm{O}$ & $\to$ & $\mathrm{H}_{2}\mathrm{O}_{2}+\mathrm{O}^{17}\mathrm{O}$ & $2.0\times0.5\times3.0\times10^{-13}\exp(460/T)\times0.9926$ & a, b\\
R236 & $\mathrm{HO}_{2}+\mathrm{HO}^{17}\mathrm{O}+\mathrm{M}$ & $\to$ & $\mathrm{H}_{2}\mathrm{O}^{17}\mathrm{O}+\mathrm{O}_{2}+\mathrm{M}$ & $2.0\times0.5\times4.2\times10^{-33}\exp(920/T)\times0.9926$ & a, b\\
R237 & $\mathrm{HO}_{2}+\mathrm{HO}^{17}\mathrm{O}+\mathrm{M}$ & $\to$ & $\mathrm{H}_{2}\mathrm{O}_{2}+\mathrm{O}^{17}\mathrm{O}+\mathrm{M}$ & $2.0\times0.5\times4.2\times10^{-33}\exp(920/T)\times0.9926$ & a, b\\
R238 & $\mathrm{C}^{17}\mathrm{O}+\mathrm{OH}+\mathrm{M}$ & $\to$ & $\mathrm{CO}^{17}\mathrm{O}+\mathrm{H}+\mathrm{M}$ & $k_{0}=1.5\times10^{-13}(300/T)^{-0.6}\times0.9935$, $k_{\infty}=2.1\times 10^{9} (300/T)^{-6.1}\times0.9935$ & a, b\\
R239 & $\mathrm{CO}+{}^{17}\mathrm{OH}+\mathrm{M}$ & $\to$ & $\mathrm{CO}^{17}\mathrm{O}+\mathrm{H}+\mathrm{M}$ & $k_{0}=1.5\times10^{-13}(300/T)^{-0.6}\times0.9826$, $k_{\infty}=2.1\times 10^{9} (300/T)^{-6.1}\times0.9826$ & a, b\\
R240 & $\mathrm{C}^{17}\mathrm{O}+\mathrm{OH}+\mathrm{M}$ & $\to$ & $\mathrm{HOC}^{17}\mathrm{O}+\mathrm{M}$ & $k_{0}=5.9\times10^{-33}(300/T)^{1.4}\times0.9935$, $k_{\infty}=1.1\times 10^{-12} (300/T)^{-1.3}\times0.9935$ & a, b\\
R241 & $\mathrm{CO}+{}^{17}\mathrm{OH}+\mathrm{M}$ & $\to$ & $\mathrm{H}^{17}\mathrm{OCO}+\mathrm{M}$ & $k_{0}=5.9\times10^{-33}(300/T)^{1.4}\times0.9826$, $k_{\infty}=1.1\times 10^{-12} (300/T)^{-1.3}\times0.9826$ & a, b\\
R242 & $\mathrm{HOC}^{17}\mathrm{O}+\mathrm{O}_{2}$ & $\to$ & $\mathrm{HO}_{2}+\mathrm{CO}^{17}\mathrm{O}$ & $2.0\times10^{-12}\times0.9955$ & a, b\\
R243 & $\mathrm{H}^{17}\mathrm{OCO}+\mathrm{O}_{2}$ & $\to$ & $\mathrm{HO}^{17}\mathrm{O}+\mathrm{CO}_{2}$ & $2.0\times10^{-12}\times0.9955$ & a, b\\
R244 & $\mathrm{HOCO}+\mathrm{O}^{17}\mathrm{O}$ & $\to$ & $\mathrm{HO}_{2}+\mathrm{CO}^{17}\mathrm{O}$ & $0.5\times2.0\times10^{-12}\times0.9911$ & a, b\\
R245 & $\mathrm{HOCO}+\mathrm{O}^{17}\mathrm{O}$ & $\to$ & $\mathrm{HO}^{17}\mathrm{O}+\mathrm{CO}_{2}$ & $0.5\times2.0\times10^{-12}\times0.9911$ & a, b\\
\hline
\multicolumn{6}{l}{\footnotesize a: \citet{Chaffin2017}, b: \citet{Young2014} for the mass-dependent fractionation factors, c: \citet{Liu2021}, d: \citet{Gregory2021}}\\
 & &  & & & \\
\end{longtable}

\end{document}